\documentclass[a4paper,11pt]{article}
\usepackage{jheppub}
\usepackage{amsmath}
\usepackage{amssymb}
\usepackage{bbold}
\usepackage{slashed}
\usepackage{booktabs}
\usepackage{physics}
\usepackage{float}
\usepackage{tabularx}
\usepackage{graphicx}
\usepackage{orcidlink}
\usepackage[dvipsnames]{xcolor}
\colorlet{forestgreen}{ForestGreen}\usepackage[normalem]{ulem}
\usepackage[dvipsnames]{xcolor}
\usepackage[normalem]{ulem}
\let\oldcite\cite
\renewcommand{\cite}[1]{\mbox{\oldcite{#1}}}

\def\beq{\begin{equation}}
\def\eeq{\end{equation}}

\title{Two-loop renormalisation and Higgs phenomenology in the five-dimensional MSSM}

\author[a,b]{Ammar~Abdalgabar,\,\orcidlink{0000-0001-9933-7313}}
\affiliation[a]{University of Hafr Al Batin, College of Science, department of physics, Hafr Al Batin 39524,
Kingdom of Saudi Arabia}
\affiliation[b]{Department of Physics, Sudan University of Science and Technology, Khartoum 407, Sudan}
\author[c,d]{Alan~S.~Cornell,\,\orcidlink{0000-0003-1896-4628}}
\affiliation[c]{Department of Physics, University of Johannesburg,
PO Box 524, Auckland Park 2006, South Africa.}
\affiliation[d]{Department of Physics, De La Salle University, 2401 Taft Avenue, Manila, 1004 Philippines}
\emailAdd{amari@uhb.edu.sa}
\emailAdd{acornell@uj.ac.za}
\author[e,c]{Aldo~Deandrea,\,\orcidlink{0000-0001-7414-6828}}
\affiliation[e]{Université Lyon 1, CNRS, IP2I, UMR 5822, Villeurbanne, France}
\emailAdd{deandrea@ip2i.in2p3.fr}
\author[f]{Howeida~ M.~Esmaeil\,\orcidlink{0009-0007-8460-4483}}
\affiliation[f]{Sudan University of Science and Technology, College of Graduate Studies, Department of
physics, Khartoum 407, Sudan}
\emailAdd{hwdmohamed@gmail.com}
\author[e,g]{Mohammed~Omer~Khojali\,\orcidlink{0000-0002-0702-262X}}
\affiliation[g]{Department of Physics, University of Khartoum, PO Box 321, Khartoum
11115, Sudan.}
\emailAdd{m.khojali@ip2i.in2p3.fr}

\abstract{
The five-dimensional Minimal Supersymmetric Standard Model (5D MSSM) provides an attractive framework in which power-law renormalisation group evolution naturally generates a sizeable trilinear stop coupling, allowing the observed Higgs boson mass to be reproduced without requiring too heavy a supersymmetric spectra. In this work we present a comprehensive two-loop analysis of the 5D MSSM, deriving the complete renormalisation group equations governing the gauge, Yukawa and soft supersymmetry-breaking sectors and examining the perturbative consistency of the theory. Particular attention is given to the ultraviolet behaviour of the model, where we demonstrate that the potentially dangerous leading two-loop contributions cancel as a consequence of the underlying $\mathcal{N}=2$ supersymmetry, leaving a well-behaved perturbative expansion. Building upon this framework, we compute the radiatively corrected Higgs effective potential including the complete Kaluza-Klein (KK) contributions through KK resummation, and investigate the resulting implications for the Higgs sector. The phenomenological viability of the model is subsequently explored through detailed numerical studies of the Higgs boson mass, electroweak precision observables, Higgs coupling modifiers and signal strengths, allowing constraints on the compactification scale and supersymmetric parameter space to be established. We find that the inclusion of the two-loop corrections preserves the characteristic power-law behaviour of the five-dimensional theory while yielding a phenomenologically viable parameter space consistent with current collider and precision measurements. Furthermore, as the KK tower provides the dominant correction to the Higgs quartic coupling, raising $m_h$ to 125 GeV, over the viable parameter space, the effective Higgs couplings remain close to their Standard Model values, showing a strong decoupling of the heavy KK modes.}

\begin{document}

\makeatletter
\def\@fpheader{}
\makeatother

\maketitle

\section{Introduction}
\label{sec:setup}

\par

The discovery of a Higgs boson with a mass of approximately $125~\mathrm{GeV}$ by the ATLAS and CMS collaborations represents one of the most significant achievements of the Large Hadron Collider (LHC), completing the particle content of the Standard Model (SM) while simultaneously providing an exceptionally sensitive probe of physics Beyond the Standard Model (BSM)~\cite{ATLAS:2012yve,CMS:2012qbp}. Since its discovery, increasingly precise measurements of the Higgs boson's production mechanisms, decay channels, and couplings have shown remarkable agreement with SM predictions, whilst still allowing room for small deviations arising from new physics at higher energy scales~\cite{ATLAS:2015yey}. Consequently, any viable extension of the SM must simultaneously reproduce the observed Higgs mass and show consistency with the growing body of precision Higgs measurements. From a theoretical perspective based on symmetries, supersymmetry (SUSY) remains one of the most compelling frameworks for physics BSM, providing a natural mechanism for stabilising the electroweak scale while offering a rich phenomenology accessible at the LHC. The absence to date of direct evidence for superpartners, together with increasingly stringent constraints from Higgs physics and searches for strongly and electro-weakly produced sparticles, has nevertheless placed significant restrictions on the viable supersymmetric parameter space. In particular, searches for third-generation squarks have established stringent lower bounds on stop masses over broad regions of parameter space, with the precise limits depending upon the assumed decay modes and supersymmetry-breaking scenario~\cite{ATLAS:2018xls}. These constraints, together with the observed Higgs boson mass and increasingly precise Higgs measurements, provide important benchmarks for assessing the viability of supersymmetric extensions of the SM.

Within this context, the Minimal Supersymmetric Standard Model (MSSM) and its extensions continue to provide a canonical framework for investigating low-energy SUSY, even if the available parameter space is in most cases constrained because the simplest weak-scale SUSY models are bounded by the LHC, Higgs physics, flavour, and dark-matter data, see for example \cite{Arbey:2012bp,Kowalska:2013ica,Feng:2013pwa,Arbey:2013fqa,Ambrogi:2017lov,Slavich:2020zjv}. Moreover, despite decades of study, the mechanism responsible for SUSY breaking remains one of the principal unresolved questions in supersymmetric model building~\cite{McGarrie:2010kh,Kribs:2013lua}. Considerable effort has therefore been devoted to constructing extensions of the MSSM attempting to simultaneously address the hierarchy problem, the supersymmetric flavour problem, and the origin of the soft SUSY-breaking sector. Higher-dimensional theories and in particular large compact extra dimensions \cite{Antoniadis:1990ew,Arkani-Hamed:1998jmv,Antoniadis:1998ig} provide an attractive alternative framework in which SUSY breaking and flavour may be addressed from a geometrical perspective, see for example \cite{Kaplan:1999ac,Kaplan:2000av}, and \cite{Sundrum:2023qsd} for a recent review. Note however that special care is necessary when dealing with Scherk-Schwarz supersymmetry breaking \cite{Branchina:2023rgi}. The simplest realisations involve a single compact extra spatial dimension, leading to a five-dimensional (5D) supersymmetric model whose low-energy description corresponds to an effective four-dimensional (4D) field theory below a finite ultraviolet cut-off. Owing to their non-renormalizable nature, gauge couplings, Yukawa couplings and soft SUSY-breaking parameters acquire an explicit dependence upon this cut-off scale, while the appearance of Kaluza-Klein (KK) excitations fundamentally modifies the renormalisation group evolution of the theory~\cite{Dienes:1998vh,Abdalgabar:2014bfa,Huang:2016dtj}.

One of the most attractive phenomenological features of the 5D MSSM is the emergence of power-law renormalisation group evolution above the compactification scale \cite{Dienes:1998vg,Barbieri:2000vh,Hall:2001pg,Ghilencea:2001bv}. For sufficiently small compactification radii, this linear running naturally generates a sizeable trilinear soft-breaking parameter $A_t$, thereby enhancing the radiative corrections to the light CP-even Higgs boson, allowing the experimentally observed Higgs mass to be reproduced without requiring excessively heavy stop squarks~\cite{Abdalgabar:2014bfa}. This feature considerably improves the naturalness of the supersymmetric spectrum and provides a motivation for studying the higher-order radiative corrections within the 5D framework. However, despite these attractive features, the perturbative consistency of higher-dimensional supersymmetric theories requires careful examination. In particular, the linear dependence of the one-loop corrections upon the ultraviolet cut-off, proportional to $\Lambda R$, naturally raises the question of whether potentially dangerous contributions proportional to $(\Lambda R)^2$ emerge at two-loop order. Demonstrating that such enhanced corrections remain under theoretical control is therefore essential if the phenomenological predictions of the 5D MSSM are to be regarded as reliable~\cite{Masip:2000yw}.

The principal objective of the present work is, therefore, to derive the complete two-loop renormalisation group equations for the 5D MSSM and to investigate their phenomenological implications for the Higgs sector. In particular, we examine the stability of the perturbative expansion, determine the evolution of the gauge, Yukawa and soft supersymmetry-breaking parameters, and assess the resulting predictions for the Higgs boson mass, the electroweak precision observables and the Higgs signal strengths. Throughout this work, emphasis is placed on establishing that the higher-dimensional theory remains both theoretically consistent and phenomenologically viable once the two-loop corrections are included. As such, the remainder of this paper is organised as follows: Section~\ref{Sec:5D MSSM} introduces the 5D MSSM, describing its field content, compactification on an $S^1/\mathbb{Z}_2$ orbifold, the resulting KK spectrum, and the boundary conditions adopted throughout the analysis. Section~\ref{Sec:Two loop RGES} presents the complete two-loop renormalisation group formalism, including the derivation of the renormalisation group equations, the effective Higgs potential, KK resummation, and wave-function renormalisation. In section~\ref{Sec:Oblique Parameters} we compute the oblique parameters $S$, $T$, and $U$ using KK-resummed vacuum polarisations, and compare them with electroweak precision constraints. In section~\ref{Sec:Higgs Phenomenology} we analyse the Higgs couplings and signal strengths, comparing our results with the existing literature and deriving constraints on the compactification scale. Section~\ref{Sec:Numerical} presents the numerical results, and finally, we give our conclusions in section~\ref{Sec:Conclusion}.

\section{Five-Dimensional MSSM}
\label{Sec:5D MSSM}

\subsection{Model construction}

\par
In the following we consider the five-dimensional Minimal Supersymmetric Standard Model (5D MSSM) compactified on the orbifold $S^1/\mathbb{Z}_2$ with compactification radius $R$. Orbifold compactifications provide one of the simplest and most phenomenologically attractive realisations of higher-dimensional supersymmetric theories, allowing the underlying 5D theory to reproduce the chiral particle content of the 4D MSSM while preserving the 4D gauge symmetry of the SM. The fifth-dimensional coordinate satisfies the orbifold identification $y\sim-y$, with
\[
y\in[-\pi R,\pi R] \; ,
\]
such that fields may be assigned definite $\mathbb{Z}_2$ parities under the orbifold symmetry. Fields with even parity possess massless zero modes that are identified with the standard MSSM particles, whereas odd fields contain only massive KK excitations.

The gauge and Higgs multiplets are bulk fields in this model and propagate throughout the 5D space-time, while the chiral fermions are localised on the orbifold fixed points (brane). These brane-localised interactions generate the supersymmetric $\mu$-term together with the Yukawa interactions responsible for the observed fermion masses after electroweak symmetry breaking. The theory reduces to the MSSM below the compactification scale, whereas above this scale the complete KK spectrum becomes dynamically relevant and fundamentally modifies the renormalisation group evolution of the theory. The field content, orbifold parities and zero modes of the theory are indicated in Table 
\ref{tab1}.

\begin{table}[htb]
\centering
\caption{The matter content for all superfields of chiral fermions on the brane alongside Higgs and gauge vector superfields in the bulk. The superscript $f=1,2,3$ denotes the family index.
\\}\label{tab1}
\renewcommand{\arraystretch}{1.2}
\begin{tabular}{@{}l c c c c c@{}}
\toprule
Superfields & Brane & Bulk & $U(1)_Y$ & $SU(2)_L$ & $SU(3)_c$ \\
\midrule
$\tilde q^{f}$ & \checkmark & -- & $1/6$  & $\mathbf{2}$ & $\mathbf{3}$ \\
$\tilde d^{f}$ & \checkmark & -- & $1/3$  & $\mathbf{1}$ & $\mathbf{\bar{3}}$ \\
$\tilde u^{f}$ & \checkmark & -- & $-2/3$ & $\mathbf{1}$ & $\mathbf{\bar{3}}$ \\
$\tilde l^{f}$ & \checkmark & -- & $-1/2$ & $\mathbf{2}$ & $\mathbf{1}$ \\
$\tilde e^{f}$ & \checkmark & -- & $1$    & $\mathbf{1}$ & $\mathbf{1}$ \\
\addlinespace
$\tilde H_d$   & -- & \checkmark & $-1/2$ & $\mathbf{2}$ & $\mathbf{1}$ \\
$\tilde H_u$   & -- & \checkmark & $1/2$  & $\mathbf{2}$ & $\mathbf{1}$ \\
\addlinespace
$\tilde B_V$   & -- & \checkmark & $0$    & $\mathbf{1}$ & $\mathbf{1}$ \\
$\tilde W_V$   & -- & \checkmark & $0$    & $\mathbf{3}$ & $\mathbf{1}$ \\
$\tilde G_V$   & -- & \checkmark & $0$    & $\mathbf{1}$ & $\mathbf{8}$ \\
\bottomrule
\end{tabular}
\end{table}

\subsection{Kaluza-Klein decomposition}

\par
From a 4D perspective, every bulk 5D field can be seen as an infinite tower of 4D KK states. For a bulk field with mass $M$, the corresponding KK spectrum is given by
\begin{equation}
m_n^2 = M^2 + \frac{n^2}{R^2} \; , \qquad n \in \mathbb{Z} \; .
\end{equation}
The zero mode ($n=0$) is identified with the corresponding MSSM field, while the excited KK modes possess masses proportional to the compactification scale $1/R$. As the renormalisation scale increases beyond this threshold, progressively larger numbers of KK states contribute to loop corrections. This behaviour gives rise to the characteristic power-law evolution that distinguishes higher-dimensional supersymmetric theories from their 4D counterparts and underpins many of their distinctive phenomenological features.

\begin{figure}[hb!]
\begin{center}
\includegraphics[width=0.45\textwidth]{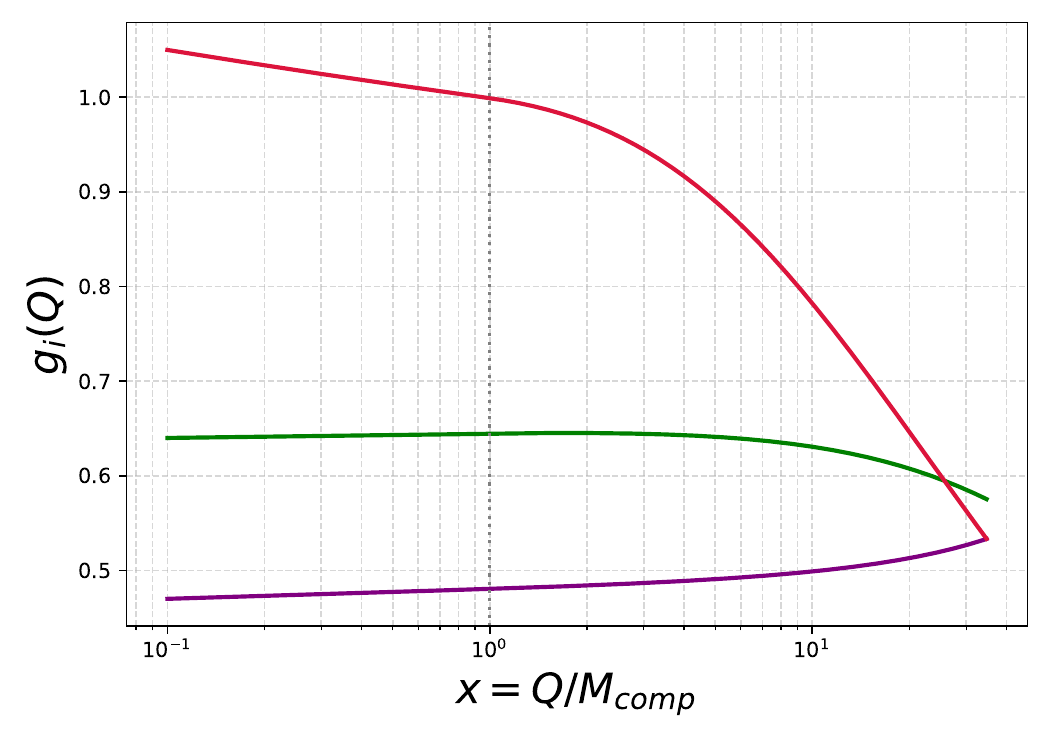}
\includegraphics[width=0.45\textwidth]{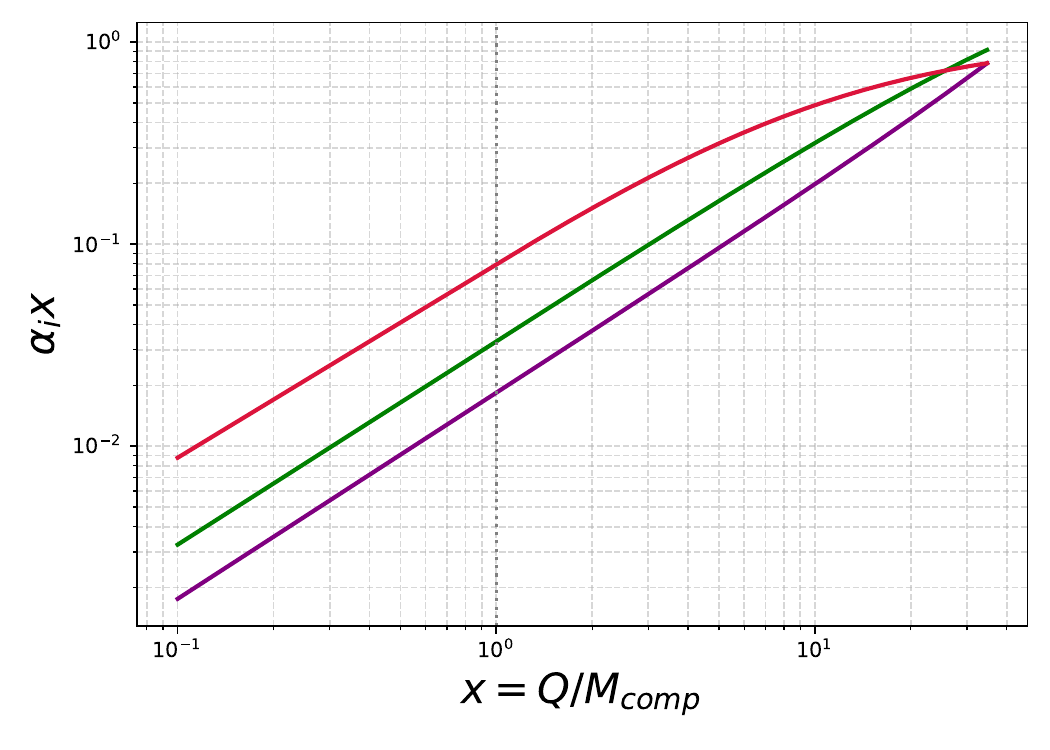}
\caption{\textbf{Left panel:} RGE evolution of the gauge couplings $g_1$ (purple), $g_2$ (green), and $g_3$ (red) as functions of $x = Q/M_{\mathrm{comp}}$. Standard logarithmic running below $x = 1$ transitions to rapid power-law evolution above the threshold due to the cumulative KK contributions, bringing the couplings toward unification near $x \approx 30$. \textbf{Right panel:} Corresponding evolution of the fine-structure constants $\alpha_i = g_i^2 / 4\pi$ on a logarithmic scale, highlighting the power-law acceleration and unification behavior at high energy scales $x \gg 1$.}
\label{fig:Gaugecoupling}
\end{center}
\end{figure}

Figure~\ref{fig:Gaugecoupling} illustrates one of the defining characteristics of the 5D MSSM; below the compactification scale, the gauge couplings evolve logarithmically according to standard 4D MSSM RGEs, but once the KK threshold is crossed ($Q > M_{\text{comp}}$), the successive excitation of higher KK levels modifies the running of the gauge couplings, both for the gauge coupling unification scale and the evolution of the soft SUSY-breaking parameters.
Since 5D gauge couplings are dimensionful, the perturbative expansion in the presence of $N_{\text{KK}} \approx Q/M_{\text{comp}}$ modes is controlled by the dimensionless effective coupling $\tilde{g}_i(Q) = g_i(Q) \sqrt{Q/M_{\text{comp}}}$ (or equivalently $\tilde{\alpha}_i(Q) = \alpha_i(Q) (Q/M_{\text{comp}})$). Plotting these rescaled couplings highlights the UV behavior and determines the natural perturbative cutoff scale $Q_{\text{cutoff}}$ of the effective field theory, where $\tilde{\alpha}_i \sim \mathcal{O}(1)$.  As will be discussed in the following sections, this enhanced running plays a central role in generating the large trilinear soft-breaking parameter $A_t$ required to reproduce the observed Higgs boson mass.

\begin{figure}[b]
\begin{center}
\includegraphics[width=0.6\textwidth]{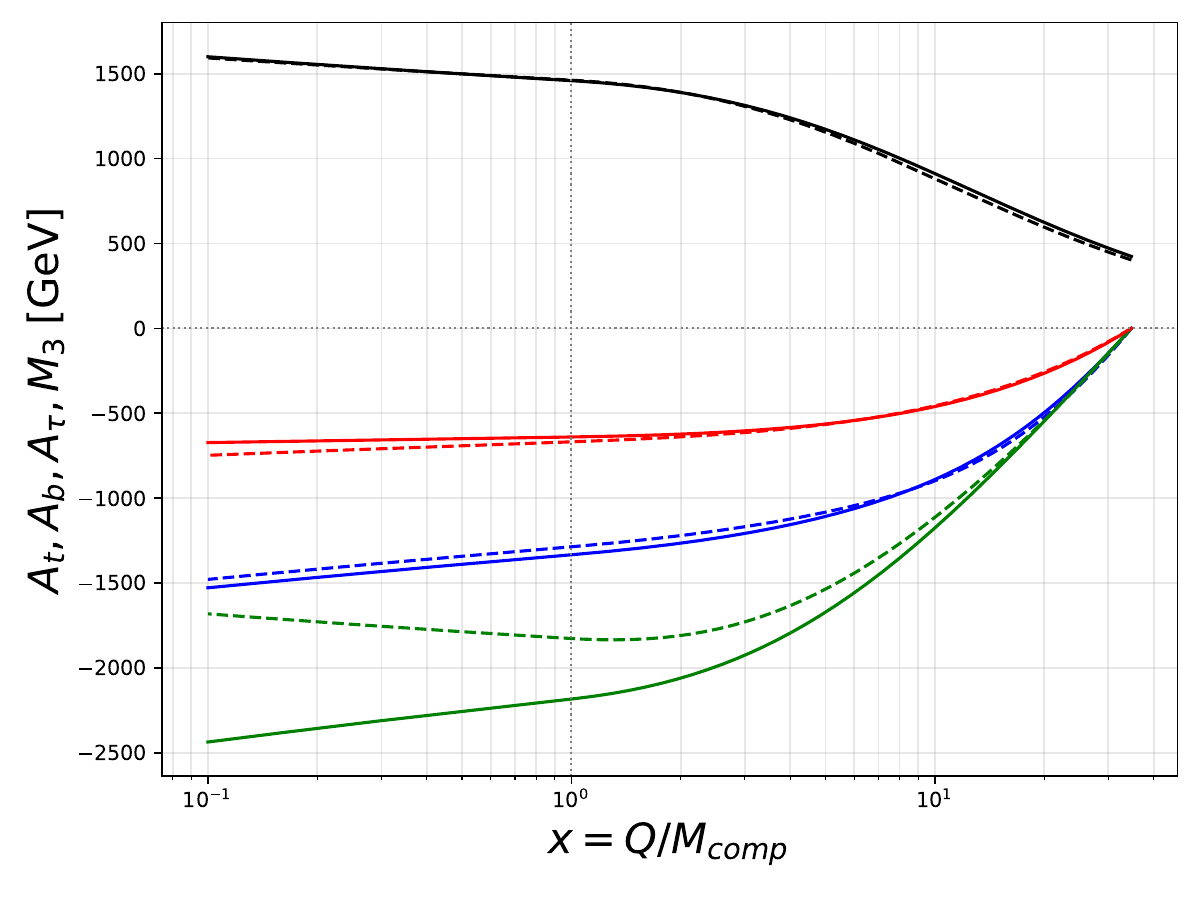}
\caption{The RGEs evolution of the trilinear scalar couplings $A_t$ (blue), $A_b$ (green), $A_\tau$ (red), and the gluino mass parameter $M_3$ (black) as a function of $x=Q/M_{\mathrm{comp}}$. Solid lines depict the one-loop 5D running, while dashed lines correspond to the two-loop 5D results. All parameters are evolved from the compactification scale at $\log_{10}(Q/\text{GeV}) \approx 5.55$, where the trilinear couplings vanish.}
\label{fig:Trilinear}
\end{center}
\end{figure}

\subsection{Boundary conditions}

Fermions and sfermions localised on the orbifold fixed points do not possess KK masses and therefore contribute only through their 4D zero modes. In contrast, every field propagating in the bulk is accompanied by an infinite tower of KK states whose collective contributions become important once the renormalisation scale exceeds the compactification threshold, $Q>1/R$. 

The mechanism responsible for SUSY breaking is not specified within the effective framework adopted here. One particularly attractive possibility is gauge-mediated supersymmetry breaking (GMSB), several realisations of which have previously been investigated in higher-dimensional models~\cite{McGarrie:2010kh}. Owing to the approximate universality of the scalar soft masses, phenomenologically viable spectra containing comparatively light squarks remain possible, even in the gaugino-mediated limit~\cite{Mirabelli:1997aj,Belanger:2015vwa}. Since the principal objective of the present work is to investigate the perturbative consistency and phenomenological implications of the complete two-loop 5D theory, we deliberately remain agnostic regarding the origin of supersymmetry breaking. Note though, an important consequence of the renormalisation group evolution in five dimensions is the natural generation of a sizeable trilinear stop coupling, $A_t$, at low energies. The enhanced stop mixing substantially increases the dominant radiative corrections to the light CP-even Higgs boson, thereby allowing the experimentally observed Higgs boson mass to be reproduced without requiring excessively heavy stop squarks~\cite{Abdalgabar:2014bfa}. This feature constitutes one of the principal phenomenological motivations for studying supersymmetric models formulated in five dimensions.

For the RGE calculation, we take into account that the SM matter superfields are localized on the brane while the gauge and Higgs superfields propagate in the bulk. For the Higgs-mass calculation, we instead assume the all-bulk realization introduced in the first scenario, since this provides the KK states necessary for a consistent evaluation of the five-dimensional effective potential. In this realization, the excited KK states provide the extra five-dimensional contribution while the zero-mode top/stop contribution reflects the standard four-dimensional top/stop correction. Therefore, the all-bulk realization utilized in the effective-potential calculation is not the same as the localization assumption used in the RGE analysis; rather, it is a conscious decision related to the Higgs-mass calculation.

Throughout this work we assume that SUSY breaking is specified at the gauge-coupling unification scale, defined by the condition $g_1=g_2=g_3$. Because of the presence of the KK spectrum, this unification scale is considerably lower than the typical one encountered in the conventional 4D MSSM, reducing the energy interval over which the soft SUSY-breaking parameters evolve. Since the subsequent low-energy phenomenology is governed primarily by the renormalisation group evolution rather than the detailed mechanism of SUSY breaking itself, the results presented here remain applicable to a broad class of SUSY-breaking scenarios. We nevertheless impose a minimal set of phenomenologically motivated boundary conditions:
\begin{itemize}
 \item[i)] We use the Yukawa and gauge couplings at the SUSY scale (1 TeV) as inputs.

 \item[ii)] We specify the value of the gluino mass, $M_3$, at 1 TeV.

 \item[iii)] We take the trilinear soft breaking terms, $A_{u/d/e}$, as vanishing at the unification scale. (see figure~\ref{fig:Trilinear}).

 \item[iv)] We specify $\mu$, $B_{\mu}$ and the sfermion soft masses at 1 TeV.
\end{itemize}
These assumptions define the boundary conditions from which the complete supersymmetric particle spectrum is determined through the coupled two-loop renormalisation group equations (RGEs) derived in the following section. Although deliberately simple, they capture the dominant features responsible for the low-energy phenomenology of the 5D MSSM while avoiding assumptions that depend on the particular ultraviolet completion of the theory.

\begin{figure}[b]
\begin{center}
\includegraphics[width=0.45\textwidth]{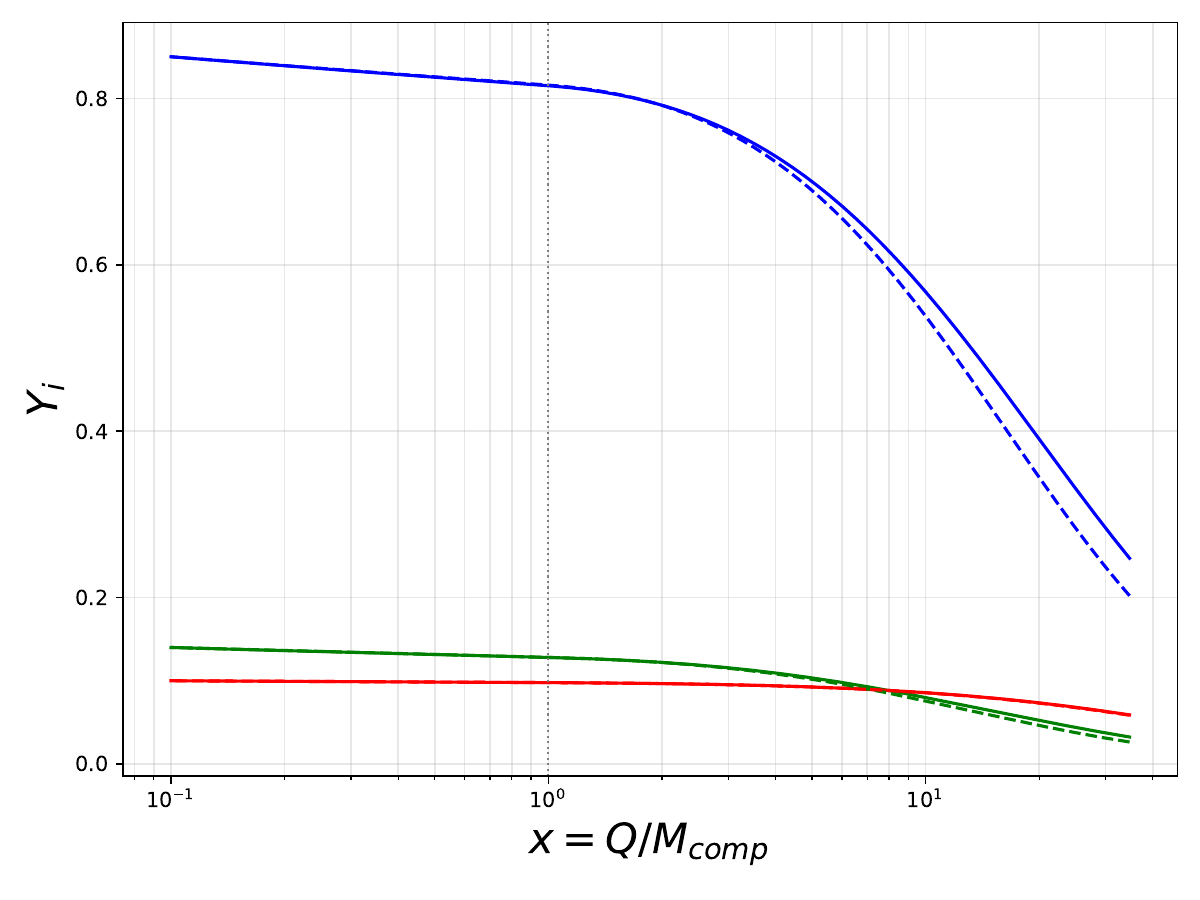}
\includegraphics[width=0.45\textwidth]{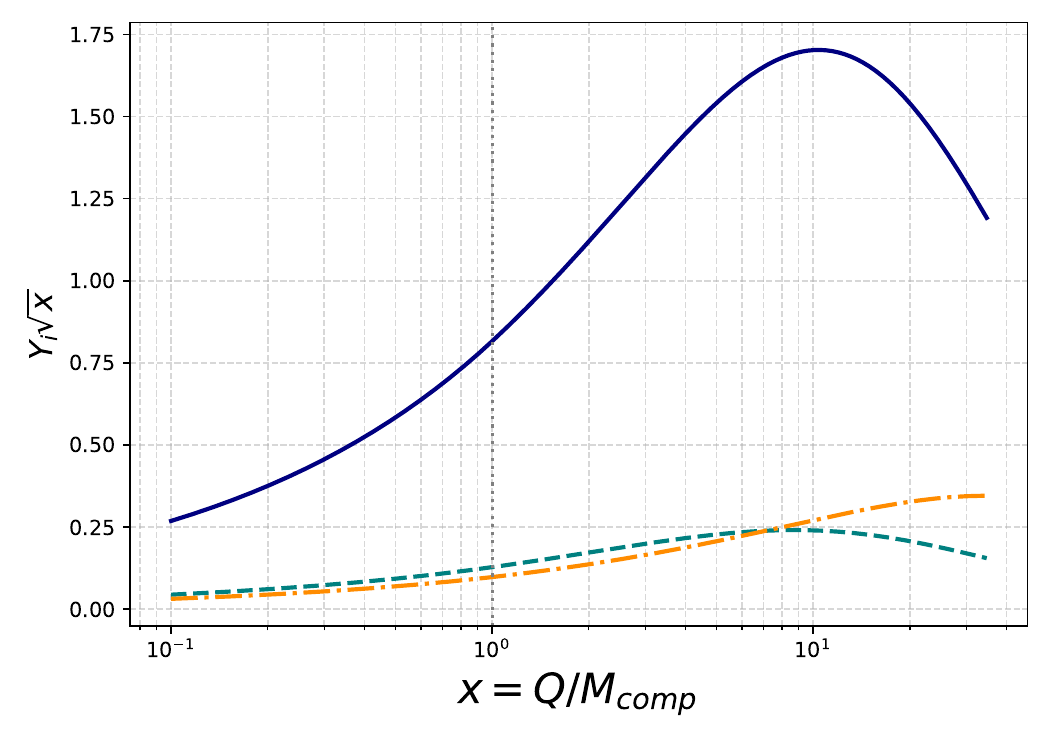}
\caption{\textbf{Left panel:} RGEs of the third-generation Yukawa couplings $Y_t$ (blue), $Y_b$ (green), and $Y_\tau$ (red) as a function of $x = Q/M_{\mathrm{comp}}$. Solid lines denote one-loop 5D running; dashed lines denote two-loop 5D running. \textbf{Right panel:} RGE of the KK-enhanced combinations $Y_t \sqrt{x}$ (solid blue), $Y_b \sqrt{x}$ (dashed green), and $Y_\tau \sqrt{x}$ (dash-dotted orange) versus $x$. The vertical dotted line indicates the compactification scale at $x = 1$.}
\label{fig:alphas5D}
\end{center}
\end{figure}


The evolution of the third-generation Yukawa couplings, shown in figure~\ref{fig:alphas5D} both as effective 4D couplings $Y_f(Q)$ (left panel) and as rescaled 5D couplings $\tilde{Y}_f(Q) \equiv Y_f(Q)\sqrt{Q/M_{\text{comp}}}$ (right panel), is of particular phenomenological importance. Above the compactification scale ($x = Q/M_{\text{comp}} > 1$), the active KK modes accelerate the RGE running. While the 4D Yukawa couplings $Y_f(Q)$ appear to decrease at higher scales due to standard negative beta-function contributions. The effective couplings $\tilde{Y}_f(Q)$, which govern loop expansions and physical interactions in the 5D theory, grow rapidly with energy due to the KK mode density scaling as $\sqrt{x}$. This KK-enhanced running of the top Yukawa coupling directly drives the evolution of the soft supersymmetry-breaking sector via the coupled RGEs, naturally generating a sizeable trilinear coupling $A_t$ over a relatively short energy range compared to standard 4D logarithmic running. Because the dominant loop corrections to the light CP-even Higgs boson mass are driven by stop masses and mixing ($X_t \approx A_t - \mu \cot\beta$), this accelerated growth provides a mechanism through which the 5D MSSM accommodates the observed 125~GeV Higgs boson mass without requiring ultra-heavy stops, thereby preserving a comparatively natural supersymmetric spectrum up to the non-perturbative cut-off scale.

\section{Two-Loop Renormalisation Group Evolution}\label{Sec:Two loop RGES}
\subsection{Renormalisation Group Equations}\label{subsec:RGEs}

A precise determination of the light CP-even Higgs boson mass requires the consistent treatment of radiative corrections over a wide range of energy scales. In supersymmetric theories this is particularly important owing to the hierarchy that generally exists between the electroweak scale and the characteristic SUSY-breaking scale, $M_{\rm SUSY}$. Fixed-order perturbative calculations therefore contain large logarithmic corrections involving ratios of these scales, together with the masses of the heavy supersymmetric particles, which can significantly reduce the convergence of the perturbative expansion if left unresolved. Within an effective field theory framework these logarithmically enhanced contributions are systematically resummed by integrating out the heavy supersymmetric degrees of freedom at the matching scale $M_{\rm SUSY}$ and subsequently evolving the effective couplings to the electroweak scale through the RGEs. The resulting low-energy parameters are then employed in the calculation of the Higgs boson mass, including the appropriate threshold corrections and radiative effects~\cite{Bahl:2019ago}.

For the 5D MSSM this procedure acquires additional importance because the renormalisation group evolution above the compactification scale is fundamentally modified by the appearance of the KK spectrum. 
Consequently, establishing the perturbative stability of the theory beyond leading order becomes an essential prerequisite for any reliable phenomenological analysis. The principal objective of this section is therefore to derive the complete set of two-loop RGEs relevant to the 5D MSSM and to demonstrate that the perturbative expansion remains well behaved despite the enhanced ultraviolet sensitivity associated with higher-dimensional theories. Particular attention is devoted to the cancellation of the leading two-loop contributions and to the subleading terms that determine the phenomenologically relevant evolution of the supersymmetric parameters.

The dominant two-loop contributions to the gauge sector are proportional to $\left(2\mu/M_c\right)^2$ and arise exclusively from two-point diagrams involving the exchange of heavy KK states~\cite{Masip:2000yw}. At first sight such terms might be expected to threaten the perturbative consistency of the effective theory owing to their enhanced ultraviolet behaviour. Remarkably, however, the underlying $\mathcal{N}=2$ SUSY of the 5D bulk enforces non-trivial cancellations amongst the leading contributions, substantially improving the ultraviolet behaviour of the theory. The relevant relations are
\begin{equation}
 P^{H}_{H} = P^{H^{'}}_{H^{'}}= 0 \; ,\; \mathrm{and}\;\; {P^{\sum^A}_{\sum^B}}= 2g^2[T(H)+C(\Sigma)]{\delta_{AB}}= g^2Q{\delta_{AB}} \; ,
\end{equation}
here $r=\delta_{AA}$ denotes the dimension of the gauge group, while $P^{H}_{H}$ and $P^{H'}_{H'}$ represent the Higgs self-energies. The relations above imply that the leading contribution to $\beta_{\tilde g}^{(2)}$ vanishes identically. This cancellation is a direct consequence of the extended $\mathcal{N}=2$ SUSY present in the bulk theory, which ensures that the one-loop counterterms are sufficient to remove the ultraviolet divergences associated with higher-order perturbative corrections. Consequently, the potentially dangerous leading two-loop contributions cancel, leaving only sub-leading corrections. These sub-leading two-loop corrections proportional to $2\mu/M_c$ originate from interactions that explicitly break the extended SUSY of the bulk, most notably the brane-localised Yukawa interactions required to generate the observed fermion masses. They therefore constitute the dominant genuine two-loop corrections governing the renormalisation group evolution of the 5D MSSM and play a central role in establishing the perturbative consistency of the theory. Following Ref.~\cite{Masip:2000yw}, the corresponding two-loop contribution to the gauge beta function may be written as:
\begin{eqnarray}
\beta_{\tilde{g}}^{(2)}&=& 2\tilde{g}^5 C(V)\{3[T(H)+T(H^{'})]+ T(f)+C(\Sigma)-9C(V)\} \nonumber \\
&&- 2g^3 r^{-1}C(H)\bigg[(4-6)g^2C(H)d(H)+\frac{1}{2}y^{Hff{'}}y_{Hff{'}}+(H \leftrightarrow H)\bigg]\nonumber\\
&&-2g^3C(\Sigma)2g^2[2T(H)-C(\Sigma)]-2g^3 r^{-1}C(f)\bigg[\frac{1}{2}y^{fkl}y_{fkl}-2g^2C(f)d(f)\bigg] \; .\label{eqn:beta2}
\end{eqnarray}
Here $d(\Phi)$ denotes the dimension of the representation associated with the field $\Phi$, while $f$ collectively denotes the chiral fermion multiplets. The various contributions in  Eq.~\eqref{eqn:beta2} may be understood directly in terms of the corresponding classes of Feynman diagrams. The first line receives contributions from diagrams containing gauge and Higgs supermultiplet particles propagating in the bulk. The overall factor of three originates from the three independent assignments of KK momentum flowing through the internal propagators, reflecting the additional combinatorial structure introduced by the compactified fifth dimension. By contrast, diagrams containing chiral fermions together with gauge supermultiplets, or gauge and adjoint supermultiplets, admit only a single KK momentum assignment and therefore do not acquire the same multiplicity factor~\cite{Di_Clemente_2002,Antoniadis_1999,Barbieri:2000vh}. This distinction between bulk and brane interactions is one of the characteristic features of orbifold compactifications and ultimately determines the structure of the surviving two-loop corrections.

For diagrams involving chiral fermions, the corresponding anomalous dimension assumes the generic form
\begin{eqnarray}
\gamma^{(2)f}_f &=& -\tilde{h}^2_f P^{f^c}_{f^c}-2g^2C(f)P^f_f+2g^4C(f)\times [2T(H)+C(\Sigma)-3C(V)] \; .
\label{eqn:gamma}
\end{eqnarray}
Eq.~\eqref{eqn:gamma} illustrates the interplay between the Yukawa and gauge sectors at two-loop order. The first term represents the Yukawa-induced contribution to the anomalous dimension, while the remaining terms encode the corresponding gauge corrections weighted by the appropriate group-theoretical factors. Although these corrections are formally sub-leading, their cumulative effect on the running of the soft SUSY-breaking parameters becomes phenomenologically significant owing to the enhanced evolution associated with the KK spectrum.

In particular, the evolution of the trilinear soft-breaking parameters and scalar masses depends sensitively upon these anomalous dimensions. Consequently, a consistent determination of the supersymmetric particle spectrum and the Higgs sector requires the complete set of coupled two-loop RGEs rather than isolated corrections to individual couplings. For completeness, the entire set of two-loop renormalisation group equations governing the gauge couplings, Yukawa couplings, gaugino masses, anomalous dimensions, trilinear soft-breaking parameters and scalar soft masses are collected in Appendix~\ref{RGES5D}. As such, throughout the main text we concentrate on the physical interpretation of the resulting evolution, while the full expressions required for numerical implementation and independent verification are provided in Appendix~\ref{RGES5D}.

\subsection{Effective Potential}
\label{Sec:EffectivePotential}
Having established the complete two-loop renormalisation group evolution, we now turn to the calculation of the Higgs effective potential. The effective potential provides a particularly powerful framework for incorporating radiative corrections into the Higgs sector, since it resums the dominant quantum effects arising from particles whose masses depend upon the Higgs vacuum expectation value. Within supersymmetric theories these radiative corrections are essential for obtaining a realistic prediction of the light CP-even Higgs boson mass, which receives substantial contributions from the top-stop sector. At one-loop order the effective potential is given by the Coleman-Weinberg expression~\cite{Coleman:1973jx}
\begin{equation}
V_{\rm CW}(H) = \frac{1}{64\pi^2} \sum_i (-1)^{2s_i} g_i \, m_i^4(H)
\left[
\ln\left(\frac{m_i^2(H)}{Q^2}\right)-c_i
\right] \; ,
\end{equation}
where $s_i$ denotes the particle spin, $g_i$ counts the corresponding degrees of freedom, $Q$ is the renormalisation scale, and the constants $c_i$ depend upon the particle species and renormalisation scheme. The alternating factor $(-1)^{2s_i}$ ensures the correct relative sign between bosonic and fermionic quantum fluctuations, reflecting the underlying supersymmetric structure of the theory.

In the 5D MSSM every KK excitation contributes an analogous Coleman-Weinberg term, requiring a summation over the complete KK spectrum~\cite{Delgado:1998qr,Barbieri:2000vh,Kubo:2001ie,Quiros:2003gg,DiClemente:2001ge,DiClemente:2002rk,Bhattacharyya:2007te}. The resulting KK resummation plays an important role in determining the radiative corrections to the Higgs sector.
The total effective potential for the neutral CP-even Higgs field $H_u^0$ may therefore be written as
\begin{equation}
V_{\rm eff}(H) = V_{\rm tree}(H) + V_{\rm soft}(H) + \sum_{\text{fields}} V_{\rm KK}(H) \; ,
\end{equation}
where the summation extends over all particle species whose masses depend upon the Higgs background field, including the top quark, stop squarks, electroweak gauge bosons, Higgs bosons, Higgsinos and their associated KK excitations. In the 5D theory the contribution of each field is replaced by the corresponding KK-resummed expression, ensuring that the radiative effects of the complete tower of massive states are consistently incorporated into the effective potential.

The tree-level contribution originates from the supersymmetric $D$-term scalar potential and is given by
\begin{equation}
V_{\rm tree}(H) = \frac{g_1^2 + g_2^2}{8} \cos^2(2\beta)
\left(
H^2-v^2\cos^2\beta
\right)^2 \; ,
\end{equation}
while the soft SUSY-breaking contribution takes the form
\begin{equation}
V_{\rm soft}(H)=\frac{m_{H_u}^2}{2}H^2 \; .
\end{equation}
We present the explicit calculation of $V_{\rm KK}(H)$ for the stop sector, including the H-expansion and KK sum with all couplings and mass dependence in the Appendix \ref{RGES5D}. With the proper substitutions, the sbottom, stau, and other sectors follow in an identical manner. We use the stop sector as the benchmark because it dominates the Higgs-mass correction.

The electroweak vacuum expectation value,
\[
v=\langle H\rangle =246.22~{\rm GeV} \; ,
\]
is determined by requiring that the effective potential be stationary at the physical vacuum. This minimisation condition, commonly referred to as the tadpole condition, fixes the Higgs soft mass parameter and ensures that electroweak symmetry breaking occurs at the correct scale. Explicitly,
\begin{equation}
\left.
\frac{dV_{\rm eff}}{dH}
\right|_{H=v}=0 \; ,
\end{equation}
thereby determining $m_{H_u}^2$ self-consistently once the complete radiative corrections have been included.

Radiative corrections also modify the couplings of the light CP-even Higgs boson to SM particles through the mixing angle $\alpha$ in the neutral CP-even Higgs sector. It is convenient to parameterise these deviations in terms of the coupling modifiers $\kappa_i$, defined as the ratio of the corresponding Higgs coupling to its SM value. At tree level these are given by
\begin{align} \label{Eq:kappa}
\kappa_V &\nonumber= \sin(\beta-\alpha) \; ,\\
\kappa_t & \nonumber= \frac{\cos\alpha}{\sin\beta} \; ,\\
\kappa_b &= -\frac{\sin\alpha}{\cos\beta} \; .
\end{align}

These coupling modifiers provide a direct connection between the theoretical predictions of the model and experimental measurements of Higgs production and decay rates. They therefore play an important role in the phenomenological analysis presented in section~\ref{Sec:Numerical}.

\subsection{KK Resummation}\label{ssec:KK Resu}
The dominant radiative corrections to the Higgs sector originate from the top-stop system. In the 5D MSSM these corrections receive contributions from the complete tower of stop KK excitations and must therefore be resummed consistently. Expanding the KK contribution to the effective potential to fourth order in the Higgs field yields the KK-resummed correction to the quartic Higgs coupling~\cite{Coleman:1973jx,Carena:1995bx,Quiros:2003gg}.
Following Refs.~\cite{Delgado:1998qr,Quiros:2003gg,vonGersdorff:2002as,Barbieri:2000vh,Kubo:2001ie}, the result may be expressed in terms of three universal KK-resummation functions\footnote{Further details of the derivation are presented in Appendix~\ref{Vkk-resummed}.}
\begin{align}
F_0(a)&=\log\Bigl(\frac{\sinh(\pi a)}{\pi a}\Bigr) \; , \\ 
F_1(a)&=\frac{\pi a \coth(\pi a)-1}{2a^2} \; , \\
F_2(a)&=\frac{\pi\coth(\pi a)}{4a^3}+\frac{\pi^2 \mathrm{csch}^2(\pi a)}{4a^2}-\frac{1}{2a^4} \; ,
\end{align}
where the dimensionless parameter
\[
a=M_{\rm stop}R
\]
measures the stop mass in units of the compactification scale.

\begin{figure}[h]
\centering
\includegraphics[width=1.02\textwidth]{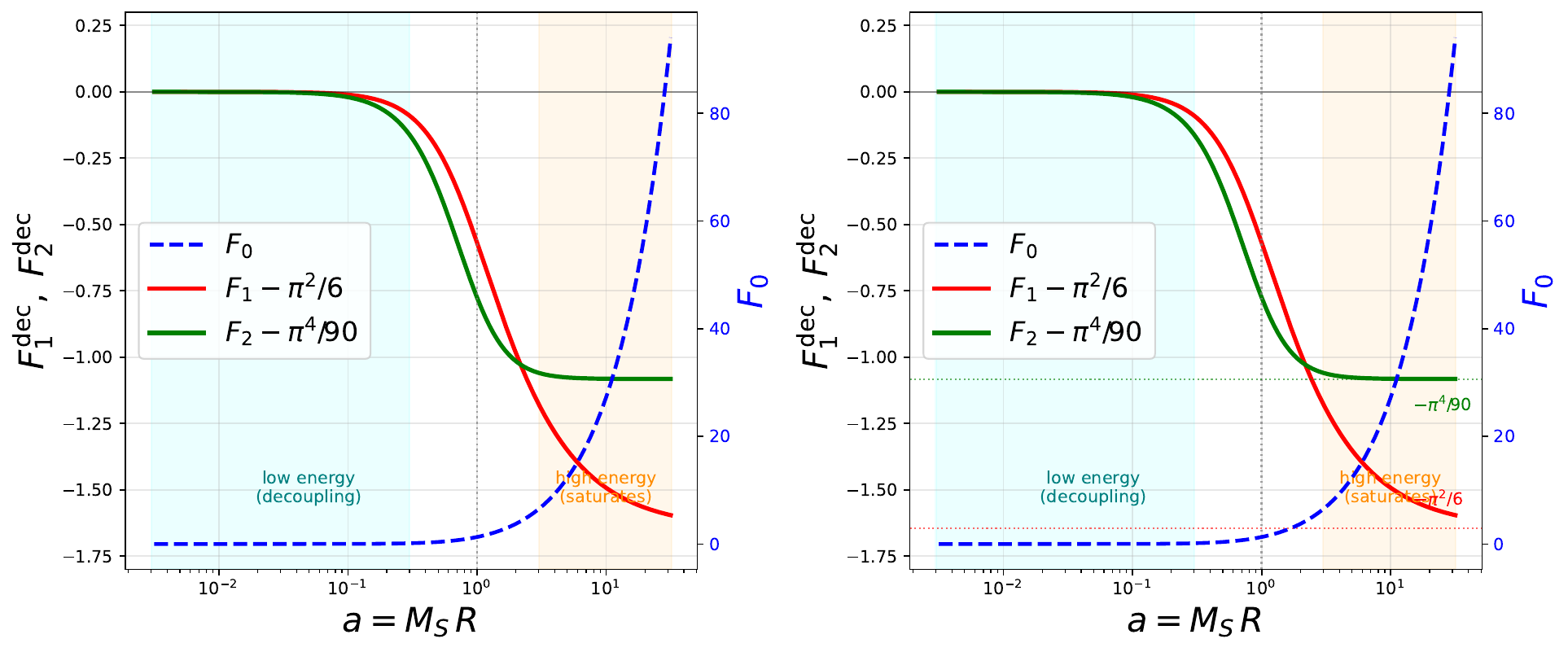}
\caption{KK threshold functions as a function of the dimensionless parameter $a=M_SR$. 
Left panel: the infrared-subtracted functions $F_1(a)-\pi^2/6$ and $F_2(a)-\pi^4/90$. 
Right panel: the same functions together with $F_0(a)$ (secondary axis). 
The cyan region corresponds to the infrared (low-energy) regime $a\ll1$, where all threshold corrections vanish and 4D physics is recovered. 
The orange region corresponds to the ultraviolet (high-energy) regime $a\gg1$, where $F_1$ and $F_2$ approach the continuum values $-\pi^2/6$ and $-\pi^4/90$, while $F_0$ grows linearly as $\pi a$. 
The vertical dotted line marks the transition scale $a\sim1$.}
\label{fig:f(a)}
\end{figure}

The functions $F_0$, $F_1$, and $F_2$ govern the KK contributions to the effective potential. In the limit $a \to 0$, they decouple, reproducing the 4D MSSM result. For finite $a$, they resum the full tower and enter the Higgs mass through the curvature of the Coleman-Weinberg potential. The transition near $a \sim 1$ (figure~\ref{fig:f(a)}) marks the onset of 5D effects in the stop sector.
\begin{equation}
\Delta\lambda_{\rm KK}
=
\frac{3y_t^4}{16\pi^2}
\left[
F_0(a)
+\frac{X_t^2}{M_S^2}F_1(a)
-\frac{X_t^4}{12M_S^4}F_2(a)
\right] \; ,\label{kk_threshold}
\end{equation}
where $M_S$ denotes the characteristic stop mass scale, $X_t=A_t-\mu/\tan\beta$ is the stop-mixing parameter, and $y_t$ is the top Yukawa coupling. Eq.~(\ref{kk_threshold}) reveals how the KK towers enhance the stop radiative corrections, directly shifting the Higgs quartic coupling. Since $m_h^2 \propto \lambda v^2$, these KK induced shifts are precisely what lift the Higgs mass into the 125 GeV range, making them the primary phenomenological signature of the 5D MSSM.

\subsection{Wave-function Renormalisation}

We determine the Higgs boson mass at zero external momentum via the derivative  of the effective potential evaluated at its minimum. However, a precise extraction of the physical pole mass requires incorporating momentum-dependent radiative corrections entering through the Higgs self-energies \cite{Hahn:2002gm, Biekotter:2017xmf}. These contributions are conventionally accounted for in the wave-function renormalisation (WFR) and constitute an essential component of precision Higgs mass computations in supersymmetric extensions 
of the SM \cite{Heinemeyer:1998np,Heinemeyer:1998kz,Borowka:2015ura}. In this work, we consistently integrate the dominant WFR corrections alongside the two-loop renormalisation group evolution outlined in the preceding subsection. This matching improves the perturbative stability of the Higgs mass prediction and significantly suppresses the residual renormalisation-scale dependence. Specifically, the Higgs WFR stems from the external momentum dependence of the one-loop self-energy, ensuring the canonical normalisation of the kinetic terms in the effective Lagrangian and yielding a gauge and scale-consistent physical pole mass \cite{Bahl:2023ead}. To this end, the relation between the bare fields $h_i^{(0)}$ and the renormalised fields $h_j$ is defined as
\begin{equation}
h_i^{(0)} = \left( \delta_{ij} + \frac{1}{2} \delta Z_{ij} \right) h_j \; ,
\end{equation}
where the field renormalisation constants $\delta Z_{ij}$ are evaluated at zero momentum transfer,
\begin{equation}
\delta Z_{ij} = -\left. \frac{\partial\Pi_{ij}(p^2)}{\partial p^2} \right|{p^2=0} \; .
\end{equation}
Here, the total self-energy $\Pi_{ij}(p^2)$ decomposes into the standard 4D contribution and the KK threshold corrections:
\begin{equation}
\Pi_{ij}(p^2) = \Pi_{ij}^{(0)}(p^2) + \Pi_{ij}^{\text{KK}}(p^2) \; .
\end{equation}
The zero-mode contribution is evaluated in terms of standard Passarino–Veltman functions, whereas the infinite KK tower is summed analytically via Poisson resummation. Upon performing the KK summation, the complete set of threshold corrections reduces to a combination of four universal functions. In particular, the tadpole contribution takes the form
\begin{equation}
\Phi(a) = \frac{2a}{\pi} \text{Li}_2\left(e^{-2\pi a}\right) + \frac{1}{\pi^2} \left[ \text{Li}_3\left(e^{-2\pi a}\right) - \zeta(3) \right] \; .
\end{equation}
The remaining contributions are encoded in the threshold functions $F_{0,1,2}$, whose explicit definitions and asymptotic properties (see figure~\ref{fig:f(a)}) are discussed in section~\ref{ssec:KK Resu}.
These four functions constitute the foundation for all KK threshold corrections, allowing the total KK contribution to the Higgs self-energy to be expressed in a compact closed form:
\begin{equation}
\Pi_{ij}^{\rm KK}(p^2)=\frac1{16\pi^2}\sum_X N_X
\left[\mathcal{T}_{ij}^{X}\Phi(a_X)+\mathcal{C}_{ij}^{X,0}F_0(a_X)+\mathcal{C}_{ij}^{X,1}F_1(a_X)
+\mathcal{C}_{ij}^{X,2}F_2(a_X)\right] \; ,
\end{equation}
where $X$ denotes the field propagating in the loop and $N_X$ accounts for its multiplicity factor. The coefficients $\mathcal{T}_{ij}^{X}$ and $\mathcal{C}_{ij}^{X,0,1,2}$ entering the master self-energy expression are collected in Table~\ref{tab:KKcoefficients}.
For fermionic contributions, the chiral coupling combinations are:
 \begin{align}
C_{ij}^{LL}&=\Gamma_j^{L*}\Gamma_i^L+\Gamma_j^{R*}\Gamma_i^R \; , \\
C_{ij}^{LR}&=\Gamma_j^{L*}\Gamma_i^R+\Gamma_j^{R*}\Gamma_i^L \; .
\end{align}
For scalar, gauge boson, Goldstone boson, and ghost loops, the couplings depend solely on the product of the two interacting vertices,
\begin{align}
S_{ij}&=\Gamma_{jSS}^{*}\Gamma_{iSS}\; , \\
V_{ij}&=\Gamma_{jVV}^{*}\Gamma_{iVV} \; , \\
P_{ij}&=\Gamma_{jGG}^{*}\Gamma_{iGG} \; , \\
G_{ij}&=\Gamma_{j\eta\eta}^{*}\Gamma_{i\eta\eta} \; .
\end{align}
The quartic scalar interaction that enters the tadpole contribution is given by
\begin{equation}
T_{ij}=\Gamma_{ijSS} \; .
\end{equation}
We calculate the WFR by differentiating the master self-energy from the external momentum squared. The threshold functions $\Phi(a)$, $F_0(a)$, $F_1(a)$, and $F_2(a)$ depend solely on the dimensionless combination $a=mR$, therefore, they are independent of $p^2$. As such, the derivative acts only on the coefficient functions that multiply the universal threshold kernels. This characteristic greatly simplifies the KK contribution to the Higgs WFR and allows all particle sectors to be handled in a single analytic framework. The decoupling of the KK states is ensured since all terms disappear as $R\to 0$. Thus, we can rewrite Eq.~(\ref{Eq:kappa}) as:
\begin{align}
\kappa_V &= \frac{\sin(\beta-\alpha)}{\sqrt{Z_h}} \; ,\\
\kappa_u &= \frac{\cos\alpha}{\sin\beta}\,\frac{1}{\sqrt{Z_h}} \; ,\\
\kappa_d &= -\frac{\sin\alpha}{\cos\beta}\,\frac{1}{\sqrt{Z_h}}\; ,
\end{align}
where $\alpha$ is the mixing angle of the CP‑even Higgs sector in the MSSM.

\begin{table}[htbp]
\centering
\caption{Coefficients of the universal KK threshold functions $\mathcal{T}_{ij}^{X}$ and $\mathcal{C}_{ij}^{X,n}$ ($n=0,1,2$) for various field sectors.
\\}
\label{tab:KKcoefficients}
\renewcommand{\arraystretch}{1.6}
\begin{tabular}{@{}c c c c c@{}}
\toprule
Sector & $\mathcal{T}_{ij}^{X}$ & $\mathcal{C}_{ij}^{X,0}$ & $\mathcal{C}_{ij}^{X,1}$ & $\mathcal{C}_{ij}^{X,2}$ \\
\midrule
Tadpole & $-T_{ij}$ & $0$ & $0$ & $0$ \\
\addlinespace
Scalar & $0$ & $S_{ij}$ & $\frac{\pi R p^2}{6 m_S} S_{ij}$ & $\frac{\pi^2 R^2 p^4}{60 m_S^2} S_{ij}$ \\
\addlinespace
Fermion & $0$ & 
  $\begin{aligned} &2 m_f^2 C_{ij}^{LR} \\ &+ (2p^2 - 8m_f^2) C_{ij}^{LL} \end{aligned}$ & 
  $\begin{aligned} &-\frac{\pi R p^2 m_f}{3} C_{ij}^{LR} \\ &-\frac{\pi R p^2}{3m_f} (p^2 - 2m_f^2) C_{ij}^{LL} \end{aligned}$ & 
  $\begin{aligned} &-\frac{\pi^2 R^2 p^4}{30} C_{ij}^{LR} \\ &-\frac{\pi^2 R^2 p^4}{30 m_f^2} (p^2 - 2m_f^2) C_{ij}^{LL} \end{aligned}$ \\
\addlinespace
Gauge boson& $0$ & $V_{ij}$ & $\frac{\pi R p^2}{6 m_V} V_{ij}$ & $\frac{\pi^2 R^2 p^4}{60 m_V^2} V_{ij}$ \\
\addlinespace
Ghost & $0$ & $G_{ij}$ & $\frac{\pi R p^2}{6 m_\eta} G_{ij}$ & $\frac{\pi^2 R^2 p^4}{60 m_\eta^2} G_{ij}$ \\
\addlinespace
Goldstone & $0$ & $P_{ij}$ & $\frac{\pi R p^2}{6 m_G} P_{ij}$ & $\frac{\pi^2 R^2 p^4}{60 m_G^2} P_{ij}$ \\
\addlinespace
Chargino & $0$ & 
  $\begin{aligned} &2 m_{\chi^{\pm}}^2 C_{ij}^{LR} \\ &+ (2p^2 - 8m_{\chi^{\pm}}^2) C_{ij}^{LL} \end{aligned}$ & 
  $\begin{aligned} &-\frac{\pi R p^2 m_{\chi^{\pm}}}{3} C_{ij}^{LR} \\ &-\frac{\pi R p^2}{3m_{\chi^{\pm}}} (p^2 - 2m_{\chi^{\pm}}^2) C_{ij}^{LL} \end{aligned}$ & 
  $\begin{aligned} &-\frac{\pi^2 R^2 p^4}{30} C_{ij}^{LR} \\ &-\frac{\pi^2 R^2 p^4}{30 m_{\chi^{\pm}}^2} (p^2 - 2m_{\chi^{\pm}}^2) C_{ij}^{LL} \end{aligned}$ \\
\addlinespace
Neutralino & $0$ & 
  $\begin{aligned} &2 m_{\chi^0}^2 C_{ij}^{LR} \\ &+ (2p^2 - 8m_{\chi^0}^2) C_{ij}^{LL} \end{aligned}$ & 
  $\begin{aligned} &-\frac{\pi R p^2 m_{\chi^0}}{3} C_{ij}^{LR} \\ &-\frac{\pi R p^2}{3m_{\chi^0}} (p^2 - 2m_{\chi^0}^2) C_{ij}^{LL} \end{aligned}$ & 
  $\begin{aligned} &-\frac{\pi^2 R^2 p^4}{30} C_{ij}^{LR} \\ &-\frac{\pi^2 R^2 p^4}{30 m_{\chi^0}^2} (p^2 - 2m_{\chi^0}^2) C_{ij}^{LL} \end{aligned}$ \\
\bottomrule
\end{tabular}
\end{table}
 
\begin{figure}[h]
\centering\includegraphics[width=0.48\textwidth]{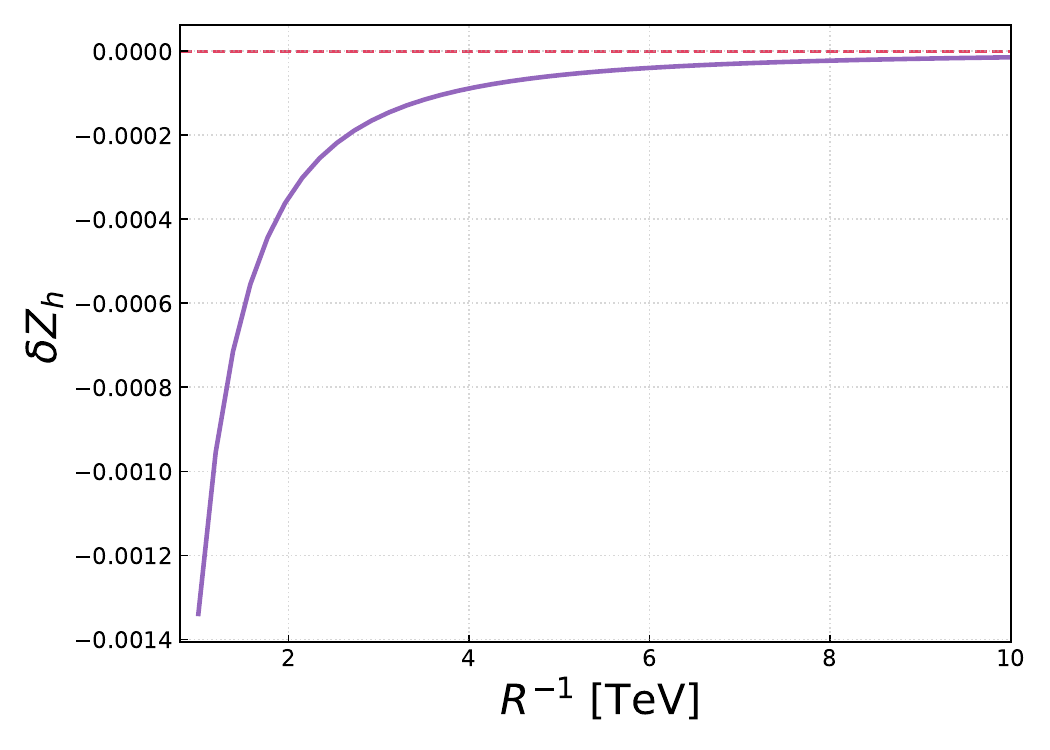}
\includegraphics[width=0.48\textwidth]{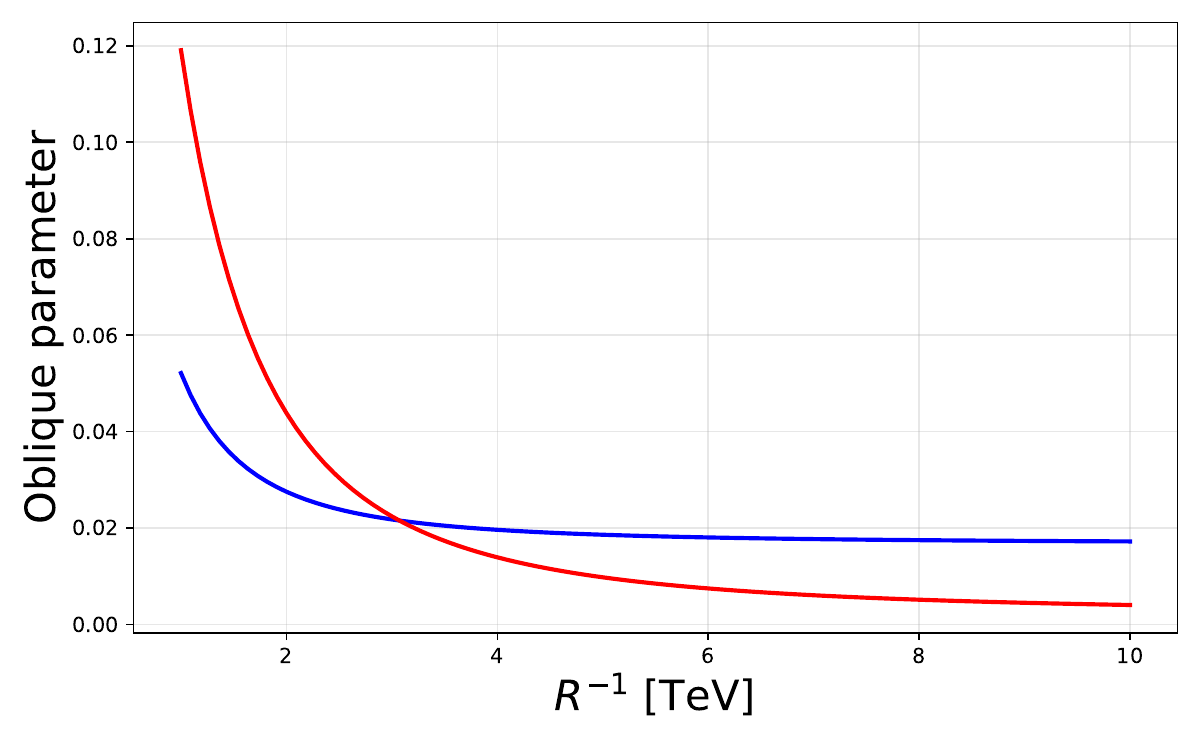}
\caption{\textbf{Left panel:} Decoupling behaviour of the Higgs WFR correction $\delta Z_h$ as a function of the compactification scale $1/R$ in the 5D MSSM. The curve is evaluated for $\tan\beta = 10$, $M_S = 1.5\text{ TeV}$, $X_t = 1.5\text{ TeV}$, and Higgsino mass parameter $\mu = 500\text{ GeV}$. The solid blue line denotes the full 5D KK contribution, while the red dashed horizontal line represents the zero baseline ($\delta Z_h = 0$). As $1/R \to \infty$, the wave-function correction exhibits rapid power-law suppression and vanishes smoothly, recovering the 4D effective field theory limit. \textbf{Right panel:} The electroweak oblique parameters $S$ (blue) and $T$ (red) as a function of the compactification scale $1/R$ (in units of \text{TeV}). The curves illustrate the expected decoupling behaviour of the 5D MSSM KK excitations, where both parameters asymptotically approach zero as the compactification scale increases.}
\label{fig:zh}
\end{figure}

\section{Electroweak Precision Constraints}\label{Sec:Oblique Parameters}

\subsection{Oblique Parameters}

Although the Higgs sector provides one of the most stringent probes of the 5D MSSM, precision electroweak observables offer an independent and highly complementary test of the model. New particles propagating in loops modify the vacuum polarisation amplitudes of the electroweak gauge bosons, leading to corrections that may be parameterised in terms of the oblique parameters $S$, $T$, and $U$. Since these quantities are constrained experimentally with high precision, they provide an important consistency check on any proposed extension of the SM. In the present framework, the complete KK tower contributes to the electroweak vacuum polarisations, consequently, the oblique parameters provide a direct probe of the compactification scale and furnish complementary constraints to those obtained from the Higgs sector and direct collider searches.

We estimate the impact of the 5D MSSM on electroweak precision observables, and evaluate the oblique parameters $S$, $T$, and $U$ in terms of the transverse gauge boson vacuum polarization functions which are given in Refs.~\cite{Peskin:1990zt,Peskin:1991sw}. 

Following the techniques developed in Ref.~\cite{Appelquist:2000nn}, the cumulative effect of the bulk field's KK towers can be written as
\begin{align}
S &= \sum_{n=1}^{\infty} S^{(n)} \; , \qquad
T = \sum_{n=1}^{\infty} T^{(n)} \; , \qquad
U = \sum_{n=1}^{\infty} U^{(n)} \; .
\end{align}
The one-loop vacuum polarisation induced by a bulk field with a zero-mode mass $m_i$ can be evaluated in closed form via Poisson resummation (see Appendix~\ref{oblique}). To determine the oblique parameters, only the vacuum polarisations and their first derivatives evaluated at $q^2=0$ are required:
\begin{align}
\Pi'_{XY}(0) &= \sum_i \frac{C_i^{XY}}{96\pi} \frac{R}{m_i} \coth(\pi R m_i) \; , \\[6pt]
\Pi_{XY}(0)   &= \sum_i \frac{C_i^{XY}}{16\pi^2} \log\!\Bigl[\sinh^2(\pi R m_i)\Bigr] + \text{const.} \; ,
\end{align}
where the additive constant cancels the differences that characterise $T$. The expressions above capture the full 5D contribution, including that of the zero modes. The corresponding gauge and flavour quantum numbers are encoded in the coefficients $C_i^{XY}$ (Table~\ref{C:XY}):
\begin{equation}
C_i^{XY} = N_c\; \kappa_i\; \sum_{\text{helicities}} (T^X_i)(T^Y_i) \; ,
\end{equation}
with $\kappa_i = 1$ for fermions, $\kappa_i = \tfrac{1}{2}$ for complex scalars (under SUSY normalisation), and $N_c$ denoting the colour factor ($N_c = 3$ for quarks, $N_c = 1$ for leptons). Here, $1$ corresponds to the charged $W$ boson ($T_3 = +1/2$), $3$ to the neutral $Z$ boson component ($T_3$), and $Q$ to the electromagnetic current. The indices $X$, $Y$ run over $\{1, 3, Q\}$. A detailed derivation of the KK-resummed formulas is provided in Appendix~\ref{oblique}. 

For fields present in the SM, namely the fermions, gauge bosons, and Higgs doublet, the zero-modes ($n = 0$) are identified with the SM particles themselves. Consequently, new physics contributions from the extra dimension arise exclusively from the non-zero KK excitations ($n \ge 1$). Subtracting the SM zero-mode contribution yields:
\begin{align}
\Pi'_{XY}(0)_{\text{new}} &= \Pi'_{XY}(0)_{\text{5D}} - \frac{C_i^{XY}}{96\pi^2 m_i^2} \; , \\[6pt]
\Pi_{XY}(0)_{\text{new}}   &= \Pi_{XY}(0)_{\text{5D}} - \frac{C_i^{XY}}{16\pi^2}\log(m_i^2) \; .
\end{align}
For new supersymmetric states without SM counterparts, such as neutralinos, charginos, and sfermions, the full 5D contribution, including the zero mode, represents entirely new physics. Accordingly, no SM zero-mode subtraction is performed for these fields.

These functions provide a convenient representation of the zero-mode subtraction,
\begin{equation}
    G(x) = x\coth x - 1 \; , \qquad H(x) = \log\!\left(\frac{\sinh x}{x}\right) \; ,
\end{equation}
which arise naturally from taking the difference between the 5D summation and the 4D zero mode:
\begin{align}
    \Pi'_{XY}(0)_{\text{KK}} &= \sum_i \frac{C_i^{XY}}{96\pi^2 m_i^2}\; G(\pi R m_i) \; , \\[6pt]
    \Pi_{XY}(0)_{\text{KK}}   &= \sum_i \frac{C_i^{XY}}{8\pi^2}\; H(\pi R m_i) + \text{const.}
\end{align}
The relevant coefficients $C_i^{XY}$ for each particle sector are summarised in Table~\ref{C:XY}. Here, $R^{\tilde{f}}$ denotes the sfermion mixing matrices, $U$ and $V$ diagonalise the chargino mass matrix, and $N$ diagonalises the neutralino mass matrix.\footnote{The real scalar normalisations set $\kappa_h = 1$ after the Goldstone mode is absorbed by the gauge sector. The weak isospin quantum numbers for the Higgs doublets are $T_3 = +1/2$ for $H_u$ and $T_3 = -1/2$ for $H_d$.} The Higgs sector mixing is parameterised by the angle $\alpha$.

\begin{table}[htbp]
\centering
\caption{The coefficients $C_i^{XY}$ entering the KK-resummed vacuum polarisations, where the superscripts $X,Y \in \{1, 3, Q\}$ correspond to the $W$ boson, $Z$ boson, and photon, respectively.}\label{C:XY}
\renewcommand{\arraystretch}{1.5}
\begin{tabular}{@{}lcccc@{}}
\toprule
Particle & $C^{11}$ & $C^{33}$ & $C^{3Q}$ & $C^{QQ}$ \\
\midrule
\multicolumn{5}{c}{\textbf{SM fermions}} \\
$u,c,t$ & $3\cdot\frac{1}{4}$ & $3\cdot\frac{1}{4}$ & $3\cdot\frac{1}{3}$ & $3\cdot\frac{4}{9}$ \\
$d,s,b$ & $3\cdot\frac{1}{4}$ & $3\cdot\frac{1}{4}$ & $3\cdot\left(-\frac{1}{3}\right)$ & $3\cdot\frac{1}{9}$ \\
$e,\mu,\tau$ & $1\cdot\frac{1}{4}$ & $1\cdot\frac{1}{4}$ & $1\cdot(-1)$ & $1\cdot 1$ \\
$\nu_e,\nu_\mu,\nu_\tau$ & $1\cdot\frac{1}{4}$ & $1\cdot\frac{1}{4}$ & $0$ & $0$ \\
\midrule
\multicolumn{5}{c}{\textbf{Sfermions (each generation)}} \\
Stop ($\tilde{t}_a$) & \multicolumn{4}{c}{$3\cdot\frac{1}{2} \sum_{i,j} R^{\tilde{t}}_{ai} T^{X}_{ij} T^{Y}_{ij}$} \\
Sbottom ($\tilde{b}_a$) & \multicolumn{4}{c}{$3\cdot\frac{1}{2} \sum_{i,j} R^{\tilde{b}}_{ai} T^{X}_{ij} T^{Y}_{ij}$} \\
Stau ($\tilde{\tau}_a$) & \multicolumn{4}{c}{$1\cdot\frac{1}{2} \sum_{i,j} R^{\tilde{\tau}}_{ai} T^{X}_{ij} T^{Y}_{ij}$} \\
\midrule
\multicolumn{5}{c}{\textbf{Higgs sector}} \\
$h^0,H^0$ & $0$ & $\kappa_h \bigl(\cos\alpha\, T_{H_u} + \sin\alpha\, T_{H_d}\bigr)^2$ & $0$ & $0$ \\
$H^\pm$ & $\kappa_h \cdot 2$ & $\kappa_h \cdot 2\cdot c_W^2/s_W^2$ & $\kappa_h \cdot 2\cdot c_W/s_W$ & $\kappa_h \cdot 2$ \\
\midrule
\multicolumn{5}{c}{\textbf{Gauge bosons}} \\
$W^\pm$ & $2$ & $0$ & $0$ & $0$ \\
$Z$ & $0$ & $\frac{c_W^2}{s_W^2}$ & $0$ & $0$ \\
$\gamma$ & $0$ & $0$ & $0$ & $1$ \\
\midrule
\multicolumn{5}{c}{\textbf{Charginos}} \\
$\tilde{\chi}^\pm_a$ & $|U_{a1}|^2$ & $|U_{a1}|^2 + |V_{a1}|^2$ & $|U_{a1}|^2$ & $|V_{a1}|^2$ \\
\midrule
\multicolumn{5}{c}{\textbf{Neutralinos}} \\
$\tilde{\chi}^0_a$ & $0$ & $\sum_{i=1}^4 |N_{ai}|^2 (T_3^i)^2$ & $\sum_{i=1}^4 |N_{ai}|^2 T_3^i Q_i$ & $\sum_{i=1}^4 |N_{ai}|^2 Q_i^2$ \\
\bottomrule
\end{tabular}
\end{table}

The KK resummed self energies, with zero-mode subtraction implemented for the SM fields, are inserted into the definitions above to yield the renormalised oblique parameters (representing the new physics contribution relative to the SM evaluated at a reference Higgs mass):
\begin{align}
S &= 16\pi \left( \Pi'_{33}(0)_{\text{new}} - \Pi'_{3Q}(0)_{\text{new}} \right) \; , \\[4pt]
T &= \frac{1}{\alpha_{\text{em}} m_Z^2} \left( \Pi_{11}(0)_{\text{new}} - \Pi_{33}(0)_{\text{new}} \right) \; , \\[4pt]
U &= 16\pi \left( \Pi'_{11}(0)_{\text{new}} - \Pi'_{33}(0)_{\text{new}} \right) \; ,
\end{align}
where $\alpha_{\text{em}} = 1/127.9$. 

\begin{figure}[htbp]
\centering
\includegraphics[width=0.6\textwidth]{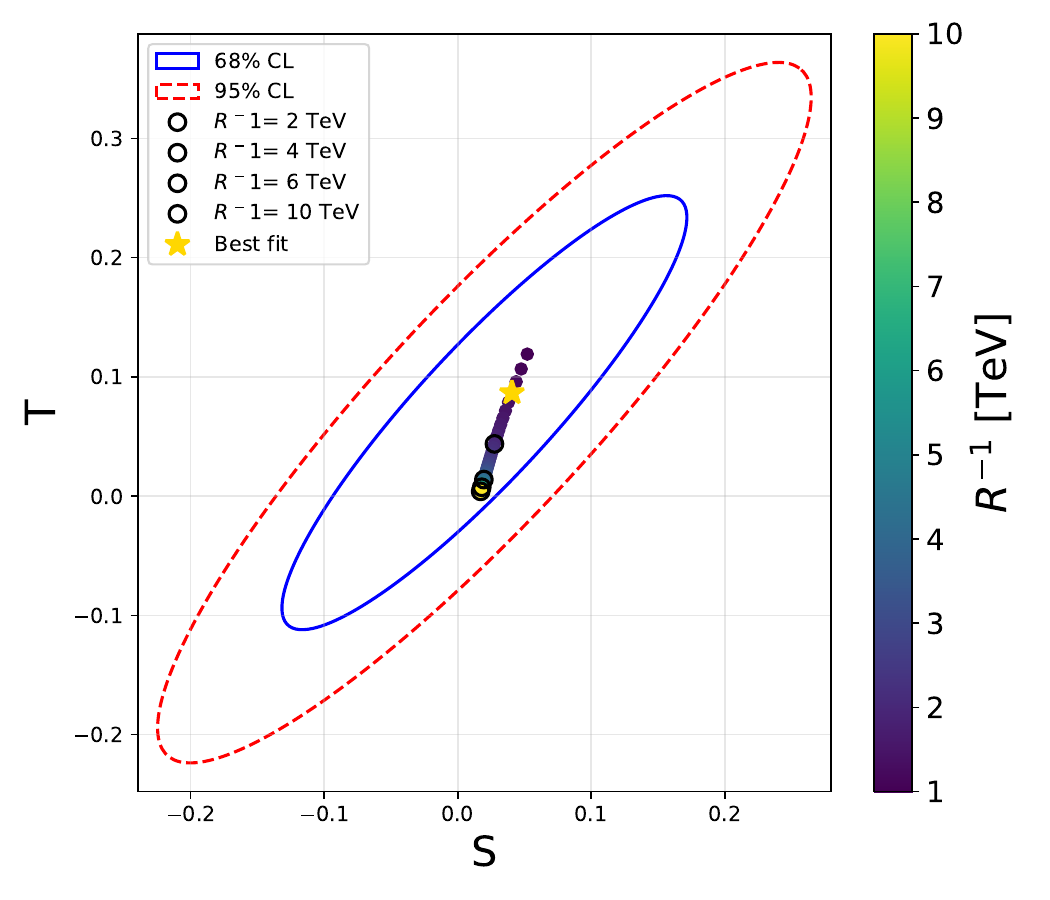}
\caption{Constraints on the $S$--$T$ plane within the 5D MSSM framework. The solid blue and dashed red ellipses represent the $68\%$ CL ($1\sigma$) and $95\%$ CL ($2\sigma$) regions from global electroweak fits, respectively. The predictions of the 5D MSSM are depicted as scatter points colour-coded by the compactification scale $1/R$ (in \text{TeV}), with specific benchmark scales ($1/R = 2, 4, 6, 10\text{ TeV}$) explicitly indicated. All model predictions lie comfortably inside the $68\%$ CL region, approaching the origin $(S, T) \approx (0, 0)$ as $1/R$ increases.}
\label{fig:ST_plane}
\end{figure}

\section{Higgs Phenomenology}
\label{Sec:Higgs Phenomenology}

\subsection{Higgs Couplings}

The discovery of a Higgs boson and the precise measurements performed at the LHC, have allowed the Higgs coupling measurements to become precision probes of new physics. Even when the masses of additional particles lie beyond the direct reach of collider experiments, their virtual effects may modify the effective Higgs couplings through radiative corrections. Consequently, the Higgs coupling modifiers provide one of the most sensitive tests of the 5D MSSM considered in this work.

In the 4D MSSM, Higgs couplings are fully determined at tree level by the mixing angles $\alpha$ and $\beta$, with loop-induced processes such as $gg \to h$ and $h \to \gamma\gamma$ receiving well-known radiative corrections. In contrast, the situation in 5D models is more fragmented with several works having analysed Higgs properties in 5D SUSY or warped scenarios. For instance, early studies of supersymmetric Higgs sectors in orbifold constructions demonstrated that the tree-level structure of the Higgs sector remains MSSM-like, with modifications arising primarily at the loop level~\cite{Delgado:1998qr}. More recent studies have focused on Higgs couplings in generic extra-dimensional setups. In warped extra dimension models, the contributions of KK fermions to the gluon fusion and di-photon amplitudes were computed, showing that the dominant corrections scale are
\begin{equation}
\kappa_{gg}, \, \kappa_{\gamma\gamma} \sim \sum_{n} \frac{v^2}{M_n^2} \; ,
\end{equation}
where $M_n$ denotes the KK mass scale~\cite{Cacciapaglia:2014rla, Azatov:2010pf}. These works demonstrated that Higgs signal strengths can deviate at the level of a few to tens of percent for KK scales in the multi-TeV range. Moreover, effective field theory approaches have been developed to study Higgs couplings in higher-dimensional theories. These formulations express loop-induced amplitudes in terms of 5D propagators and overlap integrals, capturing the non-trivial dependence on the geometry of the extra dimension~\cite{Carena:2012xa}. However, an important limitation of the existing literature is that different sectors are typically treated separately. While KK fermion contributions are well understood and MSSM loop effects (such as stops and charginos) are extensively studied in four dimensions, a complete framework combining all relevant contributions in a 5D supersymmetric setting is not fully developed.

With that in mind then, in our model, the on-shell Higgs couplings are generated from the renormalised one-particle irreducible vertices. The KK tower contribution includes Higgs WFR and vertex corrections. That is, 
\begin{equation}
\Gamma_i^{\rm ren}=\left(1+\frac12\delta Z_h\right)\left(\Gamma_i^{\rm SM}+\Delta\Gamma_i^{\rm KK}\right) \; .
\end{equation}
And therefore, the coupling modifiers can be written as
\begin{equation}
\kappa_i=\frac{\Gamma_i^{\rm ren}}{\Gamma_i^{\rm SM}},\qquad i=g,\gamma \; .
\end{equation}
In the following subsections, we explicitly evaluate the loop functions for each contributing sector, perform the regularised summation over the KK modes, and analyse the resulting deviations in the Higgs signal strengths across the 5D MSSM parameter space.

\begin{table}[htbp]
\centering
\caption{SM vs. Supersymmetric Higgs Parameter Profiles, Couplings ($\kappa$), and Extra-Dimensional Geometrical Scales.}
\label{tab:higgs_sm_susy_compact}
\small
\vskip0.1cm
\begin{tabularx}{\textwidth}{>{\raggedright\arraybackslash}p{2.3cm} >{\raggedright\arraybackslash}p{2.1cm} >{\raggedright\arraybackslash}X >{\raggedright\arraybackslash}p{2.8cm}}
\toprule
\textbf{Physics Vector} & \textbf{Parameter} & \textbf{Bound/Value (95\% CL / $1\sigma$)} & \textbf{arXiv/Ref.} \\ 
\midrule
Higgs Sector & Mass Peak ($m_h$) &
\shortstack[l]{
$125.11 \pm 0.11~\mathrm{GeV}$ (ATLAS comb. $\gamma\gamma$/$4\ell$) \\
$125.38 \pm 0.14~\mathrm{GeV}$ (CMS comb. legacy)
}
&
\shortstack[l]{
\href{https://arxiv.org/abs/1507.06712}{arXiv:1507.06712}\\
\href{https://arxiv.org/abs/2207.00043}{arXiv:2207.00043}
}
\\ \addlinespace
Higgs Couplings & Vector Gauge ($\kappa_W, \kappa_Z$) & $\kappa_W = 1.01 \pm 0.06$, $\kappa_Z = 1.02 \pm 0.06$. SM expected value: $\kappa_V = 1.00$. SUSY scales via alignment limit: $\kappa_V = \sin(\beta-\alpha) \to 1.00$. & \href{https://arxiv.org/abs/2207.00043}{arXiv:2207.00043} \newline \href{https://arxiv.org/abs/2602.23991}{arXiv:2602.23991} \\ \addlinespace
Higgs Couplings & Up-Fermion ($\kappa_t$) & $\kappa_t = 0.98 \pm 0.07$. SM expected value: $\kappa_t = 1.00$. SUSY MSSM scaling modifier parameter: $\kappa_t = \cos\alpha / \sin\beta$. & \href{https://arxiv.org/abs/2207.00043}{arXiv:2207.00043} \\ \addlinespace
Higgs Couplings & Down-Fermion ($\kappa_b, \kappa_\tau$) & $\kappa_b = 0.93 \pm 0.13$, $\kappa_\tau = 1.01 \pm 0.11$. SM: $\kappa = 1.00$. SUSY MSSM parametric enhancement: $\kappa_b = -\sin\alpha / \cos\beta \approx \tan\beta$. & \href{https://arxiv.org/abs/2207.00043}{arXiv:2207.00043} \\ \addlinespace
SUSY Higgs & Heavy Neutral ($m_A, m_H$) & Pseudoscalar $m_A \gtrsim 1.2 - 2.0 \text{ TeV}$ excluded depending on $\tan\beta$ via $\tau\tau$ decay searches. & \href{https://arxiv.org}{arXiv:2507.16400} \\ \addlinespace
SUSY Higgs & Charged States ($m_{H^\pm}$) & $m_{H^\pm} \gtrsim 800 \text{ GeV}$ bounded uniformly across top-quark association channels ($tbH^\pm$). & \href{https://arxiv.org}{arXiv:2506.06839} \\ \addlinespace
Extra-Dim Scale & Flat Radius ($R$) & Compactification scale $R^{-1} > 1.5 \text{ TeV}$ (MUED models). Spatial radius $R < 0.1\text{ mm}$ for $n=2$ (ADD models). & \href{https://worldscientific.com}{WorldSci:23500021} \newline \href{https://iop.org}{IOP:17:033015} \\
\bottomrule
\end{tabularx}
\end{table}

\subsection{Signal Strengths}

The modified Higgs couplings derived above may be translated directly into predictions for the experimentally measured signal strengths. Since these observables combine production cross sections and branching fractions, they provide a convenient means to compare theoretical predictions with the latest ATLAS and CMS measurements ~\cite{ATLAS:2016neq,ATLAS:2022vkf,CMS:2022dwd,LHCHiggsCrossSectionWorkingGroup:2013rie}. Throughout this analysis, we assume the narrow-width approximation and evaluate the signal strengths consistently using the radiatively corrected Higgs couplings obtained within the 5D MSSM.

At the LHC, the signal strengths are heavily dominated by the gluon-gluon fusion ($gg\text{F}$) channel, followed by vector-boson fusion (VBF) and associated production with vector bosons ($VH$) or top-quark pairs ($t\bar{t}H$).  Within the 5D MSSM framework, these channels exhibit a high degree of sensitivity to the compactification scale $R^{-1}$. Furthermore, the loop-induced $gg\text{F}$ production cross section, much like the $H \to \gamma\gamma$ decay amplitude, is also exceptionally sensitive to $R^{-1}$. This strong sensitivity arises because the entire infinite tower of KK excitations of bulk fields (most notably the third-generation quarks and squarks) run in the loop, providing a collective modification to the effective $Hgg$ coupling that scales with the size of the extra dimension. In contrast, tree-level production processes like VBF and $VH$ are primarily modified through the mixing of standard gauge bosons with their heavy KK counterparts, leading to corrections that scale more moderately as $\mathcal{O}(v^2 R^2)$. Consequently, the $gg\text{F}$ channel serves as the primary probe of the compactification scale in our numerical scans, while the VBF and $VH$ channels provide complementary, cleaner channels to break degeneracies in the coupling fits.

As such, the signal strength (normalised to the SM)is defined in this work as:
\begin{equation}
\mu_{gg} = \frac{\kappa_g^2\,\kappa_\gamma^2}{\kappa_H} \; .
\end{equation}
where $\kappa_{g}$ and $\kappa_{\gamma}$ encode the relative corrections to the production and decay amplitudes.

\subsection{h-$gg$ and h-$\gamma\gamma$ couplings}
For all particle spectrums, the unified formula for the on-shell KK threshold contribution can be structured as
\begin{equation}
\Delta\Gamma_{hXX}^{\rm KK}=\sum_{\phi}{\cal N}_{\phi}\left[c_0^{\phi} F^{\phi}_{0}
+c_1^{\phi} \frac{m_h^2}{4}F^{\phi}_{1}+c_2^\phi \frac{m_h^4}{16}F^{\phi}_{2}\right] \; ,
\end{equation}
where $\phi$ and $X=g,\gamma$ span the particle sectors. Table~\ref{tab:master_coefficients} summarises the universal spin-dependent coefficients and normalisation factors ${\cal N}_\phi$.

\begin{table}[tb]
\centering
\caption{Normalisation factors and universal heavy-mass expansion coefficients for the triangular form-factor contributing to $hgg$ and $h\gamma\gamma$.}\label{tab:master_coefficients}
\vskip0.1cm
\renewcommand{\arraystretch}{1.4}
\begin{tabular}{@{}c c c c c c@{}}
\toprule
Sector ($\phi$) & $c_0^\phi$ & $c_1^\phi$ & $c_2^\phi$ & $\mathcal{N}_\phi$ (Gluon) & $\mathcal{N}_\phi$ (Photon) \\
\midrule
Fermion ($f$) & $\frac{4}{3}$ & $\frac{7}{90}$ & $\frac{1}{126}$ & $N_c g_{hff}$ & $N_c Q_f^2 g_{hff}$ \\
Scalar ($S$)  & $\frac{1}{3}$ & $\frac{2}{45}$ & $\frac{1}{140}$ & $g_{hSS}$ & $N_c Q_S^2 g_{hSS}$ \\
Vector ($V$)  & $-7$ & $-\frac{11}{30}$ & $-\frac{19}{420}$ & $0$ & $N_c Q_V^2 g_{hVV}$ \\
\bottomrule
\end{tabular}
\end{table}

The coupling modifiers are, therefore, given by
\begin{equation}
\kappa_g=1+\frac12\delta Z_h+\frac{\Delta\Gamma_{hgg}^{\rm KK}}{\Gamma_{hgg}^{\rm SM}} \; ,
\end{equation}
and
\begin{equation}
\kappa_\gamma=1+\frac12\delta Z_h+\frac{\Delta\Gamma_{h\gamma\gamma}^{\rm KK}}
{\Gamma_{h\gamma\gamma}^{\rm SM}} \; .
\end{equation}
For the vector bosons, the tree-level coupling is affected by the Higgs WFR as follows:
\begin{equation}
\kappa_V=\left(1+\frac12\delta Z_h\right)\sin(\beta-\alpha) \; ,\label{eq:kappa_V}
\end{equation}
where in the decoupling limit $\sin(\beta-\alpha)\simeq 1$. The $\kappa_{u,d}$ couplings will receive both wave-function and vertex corrections:
\begin{equation}
\kappa_u=\left(1+\frac12\delta Z_h\right)\frac{\cos\alpha}{\sin\beta}+\Delta_{\rm vertex}^{u} \; ,\label{eq:kappa_u}
\end{equation}

\begin{equation}
\kappa_d=\left(1+\frac12\delta Z_h\right)\left(-\frac{\sin\alpha}{\cos\beta}\right)
+\Delta_b \; ,\label{eq:kappa_d}
\end{equation}
where $\Delta_b$ includes contributions from gluino-sbottom, chargino-stop, and KK towers.

\section{Numerical Results}
\label{Sec:Numerical}

Turning now to the numerical results, we begin by fixing $\tan \beta=10$, $X_t=2.1$ , and $M_S=1.5$ TeV, then scan over $R^{-1}$ from 1 to 50 TeV. Recall that the Higgs mass consists of the tree-level quartic coupling, a one-loop stop correction, and KK threshold corrections from Eq.~(\ref{kk_threshold}), along with the dominant two loop QCD and electroweak  contributions:
\begin{equation}
 \Delta \lambda_{total} =  \Delta \lambda_{tree}+\Delta \lambda^{(1)}_{stop}+\Delta \lambda^{(1)}_{KK}+\Delta \lambda^{(2)}_{QCD}+\Delta \lambda^{(2)}_{EW} \; ,
\end{equation}
where $m_h=\sqrt{2\Delta \lambda_{total}}v$. The KK threshold functions $F_0$, $F_1$, and $F_2$ (defined in section \ref{ssec:KK Resu}; see figure \ref{fig:f(a)}) enter the Higgs mass, WFR, the Higgs coupling modifiers, and the oblique parameters. Their dependence on $a$, that is $F_0 \sim \pi a$ for $a \gg 1$, and the constant continuum limit for $F_1, F_2$, along with all threshold corrections vanishing in the infrared ensures a smooth return to the 4D MSSM for $a \ll 1$. 

To quantify the impact on each observable, we examine the high scale behaviour of the extra-dimensional contributions. To begin, we evaluate the decoupling of the Higgs WFR correction $\delta Z_h$ as a function of the compactification scale $R^{-1}$ in the left panel of figure~\ref{fig:zh}. The benchmark scenario is fixed as described above ($\tan\beta = 10$, $M_S = 1.5\text{~TeV}$, $X_t = 1.5\text{~TeV}$) with $\mu = 500\text{~GeV}$. As illustrated, the full 5D KK contribution to $\delta Z_h$ remains strictly negative throughout the scanned range $R^{-1} \in [1, 10]\text{~TeV}$. At low compactification scales ($R^{-1} \lesssim 2\text{~TeV}$), the infinite sum over heavy KK modes yields a sizeable correction, reaching $\delta Z_h \approx -2.44 \times 10^{-3}$ near $R^{-1} = 1\text{~TeV}$. However, as $R^{-1}$ increases, the KK states become heavy and systematically decouple, leading to a rapid power law suppression of the wave function correction. Beyond $R^{-1} \approx 6\text{~TeV}$, the correction asymptotically approaches zero, cleanly restoring the decoupling limit and recovering the standard 4D effective field theory framework.

For the $S$ and $T$ parameters at one-loop order, we include the complete stop and sbottom mass matrices with mixing, the full Higgs sector ($h, H, A, H^\pm$), the charginos and neutralinos (in the higgsino approximation), and the resummed KK towers of all these fields together with the electroweak gauge bosons. When calculating the infinite KK sums we use the exact soft-cutoff functions. By adding the resulting expressions to the experimental central values, and comparing to the global electroweak fit via a $\chi^2$ constructed from the $S$, $T$ covariance matrix, we obtain the electroweak oblique parameters values as a function of the compactification scale (see the right panel of figure~\ref{fig:zh}). In this figure the KK contributions decrease monotonically as $R^{-1}$ increases, vanishing in the decompactification limit where the 4D MSSM is recovered. The $T$ parameter is more sensitive at low scales ($R^{-1} \lesssim 3$ TeV) due to KK contributions to custodial symmetry breaking, reaching $\sim 0.044$ at $R^{-1} = 2$ TeV. The $S$ parameter shows a gentler slope, remaining below 0.03 across the plotted range. For $R^{-1} \gtrsim 4$ TeV, both parameters fall below 0.015, well within the current electroweak precision bounds ($\Delta S, \Delta T \lesssim 0.05$). This confirms that the KK corrections decouple sufficiently fast to avoid tension with precision measurements.

To further assess the compatibility of the model with precision electroweak data, we examine the oblique parameters $S$ and $T$ in figure~\ref{fig:ST_plane}, which shows the parameter space predictions of the 5D MSSM projected onto the $S$--$T$ plane, alongside the experimental $68\%$ CL ($1\sigma$, solid blue ellipse) and $95\%$ CL ($2\sigma$, dashed red ellipse) contours derived from global electroweak fits. The parameter points are colour coded according to the compactification scale $R^{-1}$ spanning from $1\text{~TeV}$ to $10\text{~TeV}$. At lower compactification scales ($R^{-1} \approx 2\text{~TeV}$) the KK excitations induce noticeable corrections to the gauge boson self-energies, shifting the predictions toward slightly larger values of $T$ and $S$. As $R^{-1}$ increases, the heavy KK modes progressively decouple, leading to a smooth convergence of the model predictions toward the SM limit $(S,T) \approx (0,0)$. Crucially, across the entire scanned range of $R^{-1} \in [1, 10]\text{~TeV}$, all generated points remain comfortably within the $68\%$ CL region, demonstrating that the 5D MSSM easily evades current electroweak precision constraints without requiring fine tuned cancellations. Note that for the electroweakino sector, we include the complete chargino ($U,V$) and neutralino ($N$) mixing matrices, capturing the full MSSM mixing structure. The remaining contributions are computed at leading one-loop order, with KK corrections modelled via effective decoupling functions as in Ref.~\cite{Abdalgabar:2017cjw}. This approach illustrates the magnitude and decoupling of extra-dimensional effects without performing a full on-shell calculation, where a complete treatment with Passarino-Veltman self-energies and on-shell renormalization is deferred to a future work.

\begin{figure}[htbp]
\centering
\includegraphics[width=0.6\textwidth]{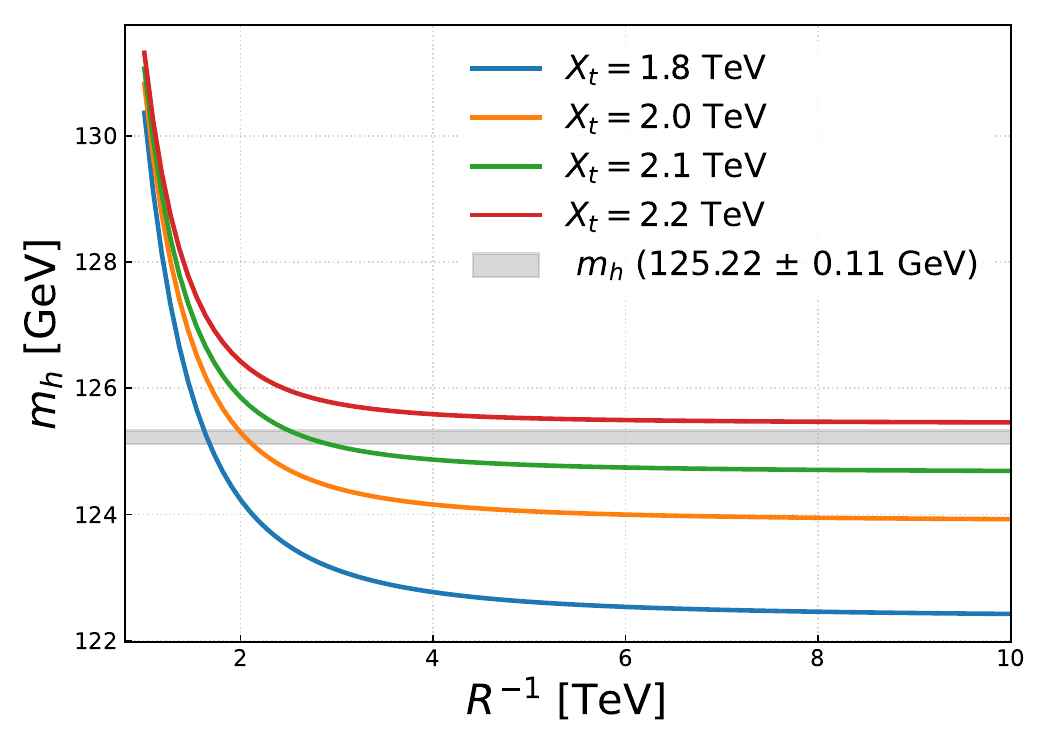}
\caption{Higgs boson mass $m_h$ as a function of the compactification scale $R^{-1}$ in the 5D MSSM, evaluated for stop mixing values $X_t \in [500, 3000]\text{ GeV}$ at fixed $M_S = 1.5\text{ TeV}$ and $\tan\beta = 10$. The coloured solid curves denote different choices of $X_t$, while the red dashed horizontal line represents the measured Higgs mass $m_h^{\text{obs}} = 125.22\text{ GeV}$ with its $\pm 1\text{ GeV}$ theoretical uncertainty band shaded in gray. As $R^{-1}$ increases, the KK threshold corrections decouple as $\mathcal{O}((1/R)^{-2})$, smoothly recovering the 4D MSSM asymptotic limit for $R^{-1} \gtrsim 10\text{ TeV}$.}
\label{fig:kk_decoupling}
\end{figure}

At high $a$ the $F_0$ is dominated by a linear growth in $a$, while $F_1$ and $F_2$ provide sub-leading shifts. At $R^{-1} = 2$ TeV, the KK correction reaches $\sim 2$ GeV, lifting $m_h$ to 125 GeV. Figure~\ref{fig:kk_decoupling} shows the KK correction to 
$m_h$ as a function of $R^{-1}=1$ TeV, where the shift reaches 6 GeV. This then drops to 25 MeV at $R^{-1} = 10$ TeV and below 1 MeV at $R^{-1} = 50$ TeV. This rapid decoupling confirms that the 5D framework reduces smoothly to the 4D MSSM at high compactification scales.

\begin{figure}[htbp]
\centering
\includegraphics[width=0.48\textwidth]{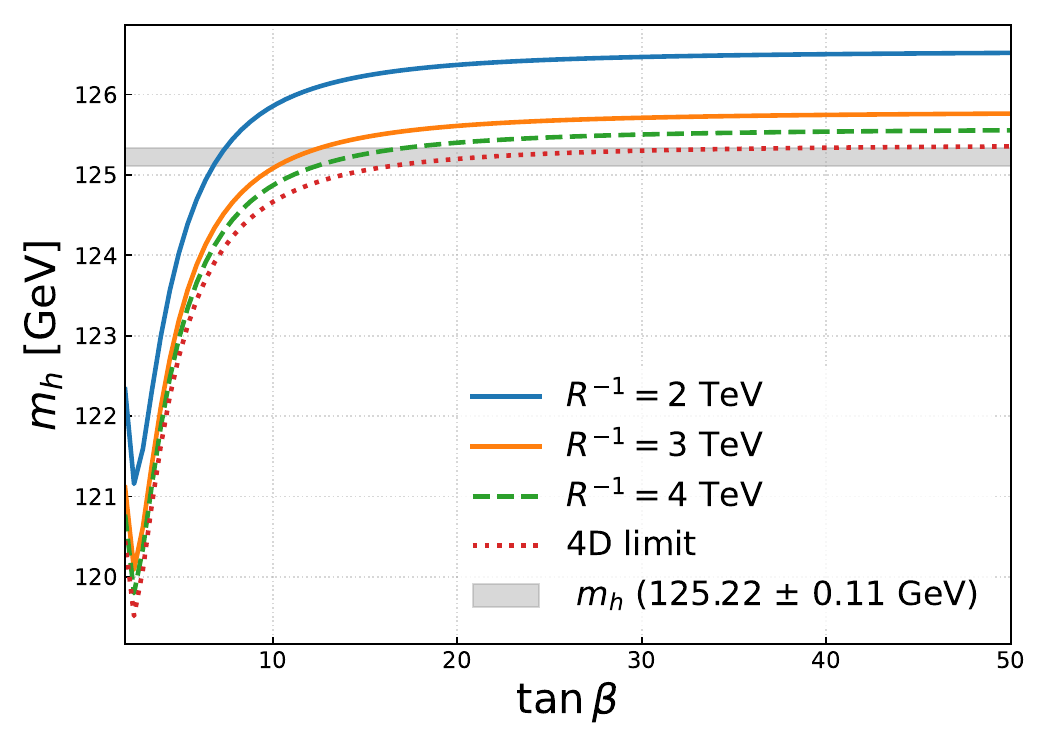}
\includegraphics[width=0.48\textwidth]{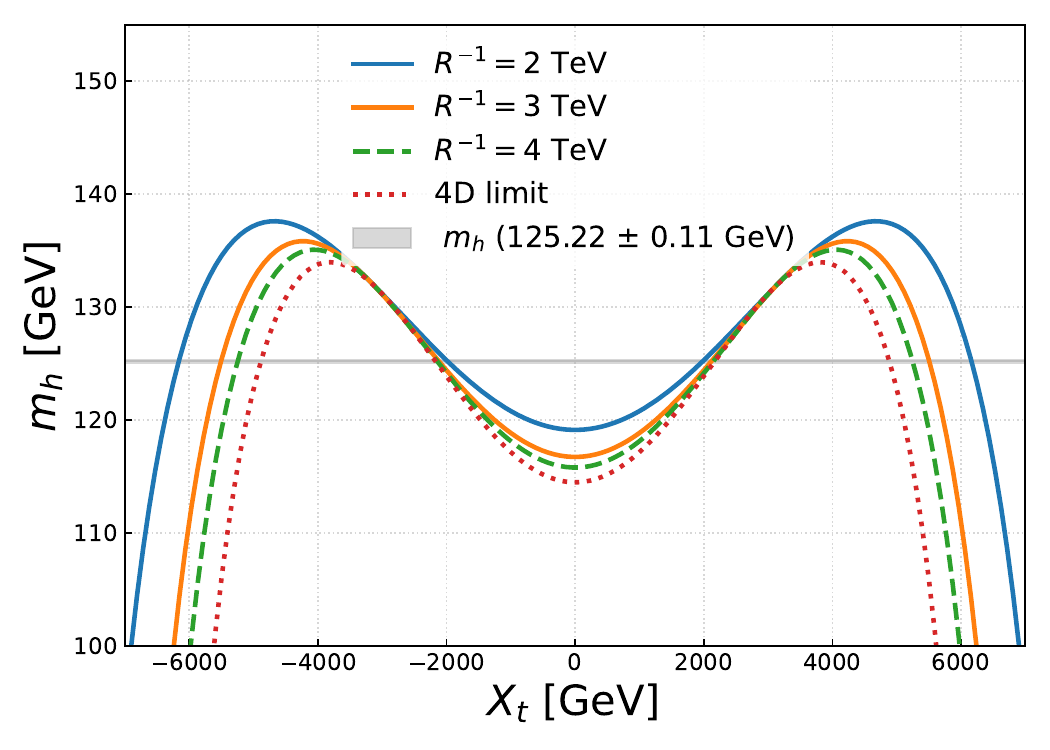}
\caption{\textbf{Left panel:} Higgs boson mass $m_h$ as a function of $\tan\beta$ for a fixed SUSY scale $M_S = 1.5\,\text{TeV}$. The dashed red curve denotes the standard 4D MSSM result, while the solid blue, orange, and green curves represent the 5D MSSM results for compactification scales $R^{-1} = 2\,\text{TeV}$, $3\,\text{TeV}$, and $4\,\text{TeV}$, respectively. The dashed gray horizontal line indicates the observed experimental value $m_h^{\text{obs}} = 125.22\,\text{GeV}$, with the gray shaded band corresponding to a $\pm 1\,\text{GeV}$ theoretical uncertainty.
\textbf{Right panel:} Higgs boson mass $m_h$ as a function of the stop mixing parameter $X_t$ for a fixed SUSY scale $M_S = 1.5\,\text{TeV}$ and ratio of vacuum expectation values $\tan\beta = 10$. The line styles and color coding for the 4D MSSM and 5D MSSM models, as well as the experimental line and theoretical uncertainty band, are identical to those in the left panel.}
\label{fig:mh_vs_tan}
\end{figure}

In figure~\ref{fig:mh_vs_tan} we show how the KK correction shifts the entire $m_h(\tan\beta)$ and $m_h(X_{t})$ curve upward by $\sim$ 2 GeV, while the shape continues to be MSSM-like, with a steep rise below $\tan \beta \sim $ 5 and a plateau above 10. Figures~\ref{fig:mh_vs_tan} show the expected behaviour similar to MSSM, that is, $m_h$ rises steeply at low $\tan \beta$ and plateaus at moderate values, then peaks near the maximal mixing ($X_t/M_S \simeq 6X_t)$. The KK correction shifts the entire curve upward, allowing $m_h=125$ GeV with a smaller $X_t$ than in the 4D MSSM.

\begin{table}[t]
\caption{5D MSSM Phenomenological viable observables.}
\label{tab:5d_mssm_master}
\small
\setlength{\tabcolsep}{3.5pt}
\renewcommand{\arraystretch}{1.25}
\begin{tabular}{@{}c c c c c c c c c c@{}}
\toprule
$R^{-1}$ (TeV) & $m_h$ (GeV) & $\delta Z_h$ & $\kappa_g$ & $\kappa_\gamma$ & $\kappa_V$ & $\kappa_u$ & $\kappa_d$ & $\kappa_h$ & $\mu_{\gamma\gamma}$ \\
\midrule
$1.0$  & $131.06$ & $-2.44 \times 10^{-3}$ & $0.99878$ & $1.00429$ & $0.99878$ & $0.99878$ & $0.99878$ & $0.99756$ & $1.01106$ \\
$2.0$  & $125.85$ & $-1.45 \times 10^{-3}$ & $0.99927$ & $1.00507$ & $0.99928$ & $0.99928$ & $0.99928$ & $0.99855$ & $1.01162$ \\
$3.0$  & $125.08$ & $-1.26 \times 10^{-3}$ & $0.99937$ & $1.00494$ & $0.99937$ & $0.99937$ & $0.99937$ & $0.99874$ & $1.01118$ \\
$4.0$  & $124.87$ & $-1.19 \times 10^{-3}$ & $0.99940$ & $1.00479$ & $0.99941$ & $0.99941$ & $0.99941$ & $0.99881$ & $1.01080$ \\
$5.0$  & $124.78$ & $-1.16 \times 10^{-3}$ & $0.99942$ & $1.00467$ & $0.99942$ & $0.99942$ & $0.99942$ & $0.99884$ & $1.01052$ \\
$7.0$  & $124.72$ & $-1.13 \times 10^{-3}$ & $0.99943$ & $1.00450$ & $0.99944$ & $0.99944$ & $0.99944$ & $0.99887$ & $1.01015$ \\
$10.0$ & $124.69$ & $-1.11 \times 10^{-3}$ & $0.99944$ & $1.00435$ & $0.99944$ & $0.99944$ & $0.99944$ & $0.99889$ & $1.00984$ \\
$50.0$ & $124.66$ & $-1.10 \times 10^{-3}$ & $0.99944$ & $1.00400$ & $0.99945$ & $0.99945$ & $0.99945$ & $0.99890$ & $1.00913$ \\
\bottomrule
\end{tabular}
\end{table}


Across the $R^{-1} \ge 1\text{ TeV}$ range, KK loop contributions to $\kappa_{g}$, $\kappa_{\gamma}$, and $\mu_{\gamma\gamma}$ induce deviations below $\sim 1\%$, preserving a smooth decoupling limit fully consistent with current and HL-LHC precision bounds \cite{LHCHiggsCrossSectionWorkingGroup:2016ypw, ATLAS:2021wqh, CMS:2018piu}. By combining the constraints from the oblique parameters $S$ and $T$, the Higgs WFR, the signal-strength modifiers $\kappa_{gg}$, $\kappa_{\gamma\gamma}$ and $\mu_{\gamma\gamma}$, and the lightest CP-even Higgs mass, the compactification scale must satisfy $M_c\gtrsim 5{-}6\,\text{TeV}$.


\section{Conclusion}\label{Sec:Conclusion}

In this paper we have derived the complete two-loop RGEs for all gauge, Yukawa, and soft SUSY breaking parameters in the 5D MSSM compactified on an $S^1/\mathbb{Z}_2$ orbifold. The KK tower drives power-law running, naturally generating $A_t \gtrsim 2~\text{TeV}$ even from vanishing trilinear boundary conditions at the unification scale. We have also verified that all couplings remain perturbative up to the cut-off. This behaviour is qualitatively consistent with the observation~\cite{Masip:2000yw} that KK-enhanced two-loop effects can become important in the Yukawa sector. In particular Ref.~\cite{Masip:2000yw} found corrections of order $\sim30\%$ in its unification-oriented analysis. Our $\sim18\%$ shifts are of comparable qualitative magnitude, although the two calculations correspond to different frameworks and boundary conditions and can not be closely compared. The size of the effect depends on the KK spectrum, the beta-function coefficients, the Yukawa couplings, and the range over which the KK-enhanced running is active. In the present analysis, $Q_{\rm high}/M_c\simeq34.7$, whereas our purpose is not to impose conventional gauge-coupling unification. Instead, we use $Q_{\rm high}\simeq3.47\times10^5$ GeV as the scale at which the gauge couplings are closest to one another and use this scale as the upper boundary for the downward RGE evolution. Thus, while the two-loop evolution changes the high-scale top Yukawa boundary value by approximately $18\%$, the resulting low-energy trilinear coupling changes by only about $3.2\%$. This demonstrates that the enhanced two-loop effects are particularly pronounced in the Yukawa sector, while the final radiatively generated trilinear coupling remains comparatively stable for this benchmark.

For the Higgs mass, we expanded the one loop KK Coleman-Weinberg potential up to $\mathcal{O}(h^4)$ for the stop sector, using analytical Poisson resummation to handle the infinite tower. By including the dominant two loop QCD and electroweak corrections, the KK threshold correction to the quartic coupling, $\Delta\lambda_{\text{KK}}$, lifts $m_h$ to $125~\text{GeV}$ without the multi-TeV stops required in the 4D MSSM.

We have also assessed the model against electroweak precision data using a phenomenological estimate of the oblique parameters $S$ and $T$. The electroweakino sector is evaluated with the complete chargino ($U, V$) and neutralino ($N$) mixing matrices, while the remaining contributions and KK corrections are computed via leading one loop expressions and effective decoupling functions. This approach illustrates the magnitude and decoupling behaviour of the extra-dimensional effects. We find that $S$ and $T$ decouple rapidly with increasing $R^{-1}$, remaining within the current $68\%$ CL global fit contours for $R^{-1} \gtrsim 1.5~\text{TeV}$. A complete calculation with Passarino-Veltman self-energies is left for future work. Finally, the Higgs coupling modifiers $\kappa_g, \kappa_\gamma, \kappa_V, \kappa_u, \kappa_d$ stay within $0.5\%$ of their SM values across most of the viable parameter space. Hence, $m_h$ is the most sensitive probe of the compactification scale, while coupling deviations will require the precision of the HL-LHC to be detected.

However, a natural extension to this work would be to include two loop electroweak threshold corrections to the Higgs potential, which could shift $m_h$ by up to $2~\text{GeV}$ and potentially close the lower end of the $R^{-1}$ window. As such, overall, the 5D MSSM with TeV scale KK towers remains a compelling and testable extension of the SM, one that will be decisively constrained at the HL-LHC.


\section*{Acknowledgements}
The work of M.O.K.~is supported by SAFE (Supporting At-Risk Researchers with Fellowships in Europe project), which is funded by the European Union (EU) under the Grant Agreement ref. 101148426. ASC was partially supported by the National Research Foundation of South Africa.
\appendix

\section{Renormalisation Group Equations for the Five-Dimensional MSSM}
\label{RGES5D}

For completeness, we collect in this appendix the complete set of two-loop renormalisation group equations employed throughout the numerical analysis presented in the main text. While only the principal features of the renormalisation group evolution are discussed in section~\ref{Sec:Two loop RGES}, the explicit expressions required for independent numerical implementation are given here. As discussed in section~\ref{Sec:Two loop RGES}, one of the defining characteristics of the 5D MSSM is the underlying $\mathcal N=2$ supersymmetry of the bulk theory. This additional supersymmetry leads to non-trivial cancellations amongst the leading two-loop contributions, ensuring that the perturbative expansion remains well behaved despite the power-law renormalisation group evolution induced by the KK spectrum. We therefore begin by summarising these cancellations before presenting the complete set of two-loop beta functions.
\subsection{Gauge Couplings in 5D at Two Loop}

We first present the complete two-loop beta functions governing the evolution of the gauge couplings.
These expressions constitute the foundation of the subsequent running of the Yukawa couplings and soft supersymmetry-breaking parameters and reproduce the characteristic power-law behaviour associated with the five-dimensional MSSM:

{\allowdisplaybreaks  \begin{align} 
\beta_{g_1}^{(2)} & =  
g_1^3\Big( \frac{199}{25}g^2_1+\frac{27}{5}g^2_2+\frac{88}{5}g^2_3 - \frac{26}{5}Y^2_u- \frac{14}{5}Y^2_d- \frac{18}{5}Y^2_e\Big)\\ 
\beta_{g_2}^{(2)} & =  
g_2^3\Big( \frac{3}{2}g^2_1+g^2_2+24g^2_3 - 6Y^2_u- 6Y^2_d- 2Y^2_e\Big)\\ 
\beta_{g_3}^{(2)} & =  
g_3^3\Big( \frac{11}{5}g^2_1+9g^2_2-40g^2_3 - 4Y^2_u- 4Y^2_d\Big) 
\end{align}} 

\subsection{Gaugino Mass Parameters in 5D at Two Loop}

The corresponding two-loop evolution equations for the gaugino mass parameters are given below.
Owing to supersymmetry, these beta functions closely mirror the structure of the gauge coupling evolution while incorporating the additional contributions arising from the trilinear soft-breaking parameters:

{\allowdisplaybreaks  \begin{align} 
\beta_{M_1}^{(2)} & =  
2g_1^2\Big( \frac{398}{25}g^2_1 M_1+\frac{27}{5}g^2_2 M_1+\frac{88}{5}g^2_3M_1+\frac{27}{5}g^2_2 M_2+\frac{88}{5}g^2_3M_3 \nonumber\\
& \ \ \ \  - \frac{26}{5}Y_u(-A_u+M_1 Y_u)  - \frac{14}{5}Y_d (-A_d+M_1 Y_d)- \frac{18}{5}Y_e(-A_e+M_1 Y_e) \Big)\\ 
\beta_{M_2}^{(2)} & =  
2g_2^2\Big( \frac{3}{2}g^2_1M_2+2g^2_2M_2+24g^2_3 M_2+\frac{3}{2}g^2_1M_1+24g^2_3 M_3 - 6Y_u (-A_u+M_2 Y_u) \nonumber\\
& \ \ \ \ - 6Y_d(-A_d+M_2 Y_d)- 2Y_e(-A_e+M_2 Y_e)\Big)\\ 
\beta_{M_3}^{(2)} & =  
2g_3^2\Big( \frac{11}{5}g^2_1M_3+9g^2_2M_3-80g^2_3M_3+\frac{11}{5}g^2_1M_1+9g^2_2M_2  \nonumber\\
& \ \ \ \ - 4Y_u (-A_u+M_3 Y_u) - 4Y_d (-A_d+M_3 Y_d)\Big) 
\end{align}}

\subsection{Anomalous Dimensions in 5D at two loop}
Furthermore, we present the corresponding two-loop evolution equations for the matter superfield anomalous dimensions.
Governed by supersymmetry, these equations closely mirror the gauge coupling evolution while incorporating two-loop wave function renormalization effects from both gauge and Yukawa interactions: 
{\allowdisplaybreaks \begin{align}  
\gamma_{\hat{q}}^{(2)} & = g_{1}^{2} \Big(\frac{8}{45} g_{3}^{2}  + \frac{1}{10} g_{2}^{2} \Big) + \frac{31}{900} g_{1}^{4}  - \frac{9}{4} g_{2}^{4}  + 8 g_{2}^{2} g_{3}^{2}  -\frac{80}{9} g_{3}^{4} + \frac{1}{2} g_{1}^{2} {Y_{u}^{\dagger}  Y_u} \nonumber \\ &\ \ \ \  -\frac{3}{2} g_{2}^{2} {Y_{u}^{\dagger}  Y_u} -2 {Y_{d}^{\dagger}  Y_d  Y_{d}^{\dagger}  Y_d} -2 {Y_{u}^{\dagger}  Y_u  Y_{u}^{\dagger}  Y_u} + \frac{1}{10} g_{1}^{2}{Y_{d}^{\dagger}  Y_d} -\frac{3}{2} g_{2}^{2}{Y_{d}^{\dagger}  Y_d} \\ 
 \gamma_{\hat{u}}^{(2)} & =  
\frac{128}{45}g_{1}^{2} g_{3}^{2}  + \frac{184}{225} g_{1}^{4} -\frac{80}{9} g_{3}^{4}  -2 {Y_u^*  Y_{d}^{T}  Y_d^*  Y_{u}^{T}} -2{Y_u^*  Y_{u}^{T}  Y_u^*  Y_{u}^{T}}+{Y_u^*  Y_{u}^{T}} \Big(3g_{2}^{2} - g_{1}^{2} \Big)\\
 \gamma_{\hat{d}}^{(2)} & =  
\frac{32}{45}g_{1}^{2} g_{3}^{2}  + \frac{34}{225} g_{1}^{4} -\frac{80}{9} g_{3}^{4}  -2 {Y_d^*  Y_{u}^{T}  Y_u^*  Y_{d}^{T}} -2{Y_d^*  Y_{d}^{T}  Y_d^*  Y_{d}^{T}}+{Y_d^*  Y_{d}^{T}} \Big(3g_{2}^{2} - \frac{1}{5}g_{1}^{2} \Big)\\
\gamma_{\hat{l}}^{(2)} & = \frac{1}{10} g_{1}^{2} g_{2}^{2} + \frac{39}{100} g_{1}^{4}  - \frac{9}{4} g_{2}^{4}-\frac{9}{10} g_{1}^{2} {Y_{e}^{\dagger}  Y_e} -\frac{3}{2} g_{2}^{2} {Y_{e}^{\dagger}  Y_e} -2 {Y_{e}^{\dagger}  Y_e  Y_{e}^{\dagger}  Y_e}\\ 
\gamma_{\hat{e}}^{(2)} & =  \frac{66}{25} g_{1}^{4}+ \frac{9}{5} g_{1}^{2} {Y_{e}^{\dagger}  Y_e}+3 g_{2}^{2} {Y_{e}^{\dagger}  Y_e} -2 {Y_{e}^{\dagger}  Y_e  Y_{e}^{\dagger}  Y_e}
\end{align} } 

\subsection{Trilinear Soft-Breaking Parameters in 5D}

The corresponding renormalization group evolution equations for the 5D trilinear soft-breaking parameters are detailed below. Reflecting the underlying supersymmetry, their beta functions share the gauge structure of the higher-dimensional theory while explicitly capturing the mixing contributions from the bulk scalar masses and gaugino soft terms:
{\allowdisplaybreaks \begin{align}  
\beta_{T_d}^{(2)} & =  
T_d \Big(\frac{167}{900} g_{1}^{4}  - \frac{9}{4} g_{2}^{4} - \frac{160}{9} g_{3}^{4}  +\frac{1}{10}g_{1}^{2} g_{2}^{2}  + \frac{8}{9} g_{1}^{2} g_{3}^{2}  +8 g_{2}^{2} g_{3}^{2} \Big) \nonumber \\
& \ \ \ \ -Y_d \Big(\frac{334}{900} g_{1}^{4}M_1  - \frac{18}{4} g_{2}^{4}M_2 - \frac{320}{9} g_{3}^{4}M_3  +\frac{1}{5}g_{1}^{2} g_{2}^{2}(M_1+M_2)\nonumber\\
& \ \ \ \  + \frac{16}{9} g_{1}^{2} g_{3}^{2}(M_1+M_3)  +16 g_{2}^{2} g_{3}^{2}(M_2+M_3)\Big)+ \frac{1}{2} g_{1}^{2} T_d {Y_{u}^{\dagger}  Y_u}-\frac{1}{2} g_{1}^{2} M_1 Y_d {Y_{u}^{\dagger}  Y_u}\nonumber\\
& \ \ \ \ -\frac{3}{2} g_{2}^{2} T_d {Y_{u}^{\dagger}  Y_u}+\frac{3}{2} g_{2}^{2}M_2Y_d {Y_{u}^{\dagger}  Y_u}-\frac{3}{10} g_{1}^{2} T_d  {Y_{d}^{\dagger}  Y_d} 
 + g_{1}^{2} Y_d T_u Y_u \nonumber\\
 & \ \ \ \ + \frac{1}{10} g_{1}^{2}M_1 Y_d {Y_{d}^{\dagger}  Y_d}+\frac{9}{2} g_{2}^{2} T_d {Y_{d}^{\dagger}  Y_d}-\frac{3}{2} g_{2}^{2}M_2 Y_d {Y_{d}^{\dagger}  Y_d}-3 g_{2}^{2}Y_d T_d Y_d -20 T_d { Y^{\dagger}_{d} Y_d  Y_{d}^{\dagger}  Y_d} \nonumber\\ & \ \ \ \ -2T_d {Y_{u}^{\dagger}  Y_u  Y_{u}^{\dagger}  Y_u}
  - 8 Y_d T_u Y_u  Y_{u}^{\dagger}  Y_u-6 T_d Y^{\dagger}_d Y_d Y_{u}^{\dagger}  Y_u-4T_u Y_u{Y_d  Y_{d}^{\dagger}  Y_d} \\ 
 \beta_{T_u}^{(2)} & =  
T_u \Big(\frac{767}{900} g_{1}^{4}  - \frac{9}{4} g_{2}^{4} - \frac{160}{9} g_{3}^{4}  +\frac{1}{10}g_{1}^{2} g_{2}^{2}  + \frac{136}{45} g_{1}^{2} g_{3}^{2}  +8 g_{2}^{2} g_{3}^{2} \Big) \nonumber \\
& \ \ \ \ -Y_u \Big(\frac{1352}{900} g_{1}^{4}M_1  - \frac{18}{4} g_{2}^{4}M_2 - \frac{320}{9} g_{3}^{4}M_3  +\frac{1}{5}g_{1}^{2} g_{2}^{2}(M_1+M_2)\nonumber\\
& \ \ \ \  + \frac{272}{45} g_{1}^{2} g_{3}^{2}(M_1+M_3)  +16 g_{2}^{2} g_{3}^{2}(M_2+M_3)\Big)- \frac{3}{2} g_{1}^{2} T_u {Y_{u}^{\dagger}  Y_u}+\frac{1}{2} g_{1}^{2} M_1 Y_u {Y_{u}^{\dagger}  Y_u}\nonumber\\
& \ \ \ \ +\frac{9}{2} g_{2}^{2} T_u {Y_{u}^{\dagger}  Y_u}-\frac{3}{2} g_{2}^{2}M_2Y_u {Y_{u}^{\dagger}  Y_u}+\frac{1}{10} g_{1}^{2} T_u  {Y_{d}^{\dagger}  Y_d} 
 +\frac{1}{5} g_{1}^{2} Y_u T_d Y_d \nonumber\\
 & \ \ \ \ - \frac{1}{10} g_{1}^{2}M_1 Y_u {Y_{d}^{\dagger}  Y_d}-\frac{3}{2} g_{2}^{2} T_u {Y_{d}^{\dagger}  Y_d}+\frac{3}{2} g_{2}^{2} Y_u {Y_{d}^{\dagger}  Y_d}-3 g_{2}^{2}Y_u T_d Y_d -20 T_u { Y^{\dagger}_{u} Y_u  Y_{u}^{\dagger}  Y_u} \nonumber\\ & \ \ \ \ -2T_u {Y_{d}^{\dagger}  Y_d  Y_{d}^{\dagger}  Y_d}
  - 8 Y_u T_d Y_d  Y_{d}^{\dagger}  Y_d-6 T_u Y^{\dagger}_u Y_u Y_{d}^{\dagger}  Y_d-4T_d Y_d{Y_u  Y_{u}^{\dagger}  Y_u}  \\
 \beta_{T_e}^{(2)} & =  
T_e \Big(\frac{303}{100} g_{1}^{4}  - \frac{9}{4} g_{2}^{4}+\frac{9}{10}g_{1}^{2} g_{2}^{2}\Big) -Y_e \Big(\frac{606}{100} g_{1}^{4}M_1  - \frac{18}{4} g_{2}^{4}M_2+\frac{18}{10}g_{1}^{2} g_{2}^{2}(M_1+M_2)\Big)\nonumber\\
& \ \ \ \ + \frac{27}{10} g_{1}^{2} T_e {Y_{e}^{\dagger}  Y_e}-\frac{9}{10} g_{1}^{2} M_1 Y_e {Y_{e}^{\dagger}  Y_e}+\frac{9}{2} g_{2}^{2} T_e {Y_{e}^{\dagger}  Y_e}-\frac{3}{2} g_{2}^{2}M_2Y_e {Y_{e}^{\dagger}  Y_e} -20 T_e { Y^{\dagger}_{e} Y_e  Y_{e}^{\dagger}  Y_e} 
\end{align} }


\subsection{One-loop KK Contribution to the Higgs Quartic Coupling}\label{Vkk-resummed}
The KK contribution is calculated in a procedure that roughly resembles the conventional MSSM stop calculation. The main distinction is that each stop's KK excitation results in its own Coleman Weinberg potential. For the $n^{th}$ KK level, the SUSY mass is
\begin{equation}
M_n^2=M_S^2+\frac{n^2}{R^2} \; ,
\end{equation}
where $M_S$ denotes the zero-mode soft stop mass and $R$ is the compactification
radius.

The field-dependent stop mass matrix reads
\begin{equation}
\mathcal{M}_{\tilde t,n}^2=
\begin{pmatrix}
M_n^2+y_t^2H^2 & y_tX_tH\\
y_tX_tH & M_n^2+y_t^2H^2
\end{pmatrix}.
\end{equation}
Introducing $a\equiv y_t^2H^2$, and $b\equiv y_tX_tH$,the eigenvalues are
\begin{equation}
m_{n,\pm}^2=M_n^2+a\pm b \; .
\end{equation}
The one-loop Coleman Weinberg potential from the complete KK tower is
\begin{equation}
V_{\rm KK}
=
\frac{3}{32\pi^2}
\sum_n
\sum_{\pm}
m_{n,\pm}^4
\left(
\ln\frac{m_{n,\pm}^2}{Q^2}
-\frac32
\right) \; .
\end{equation}
To expand around the electroweak vacuum, we define
\begin{equation}
m_{n,\pm}^2
=
M_n^2(1+\delta_{n,\pm}) \; ,\quad \delta_{n,\pm}
=
\frac{a\pm b}{M_n^2} \; ,
\end{equation}
and let 
\begin{equation}
L_n
\equiv
\ln\frac{M_n^2}{Q^2} \; .
\end{equation}
The logarithm is expanded as
\begin{equation}
\ln(1+\delta)
=
\delta
-\frac{\delta^2}{2}
+\frac{\delta^3}{3}
-\frac{\delta^4}{4}
+\mathcal{O}(\delta^5) \; ,
\end{equation}
while
\begin{equation}
(1+\delta)^2
=
1+2\delta+\delta^2 \; .
\end{equation}
We then obtain
\begin{align}
(1+\delta)^2
\left[
A
+\delta
-\frac{\delta^2}{2}
+\frac{\delta^3}{3}
-\frac{\delta^4}{4}
\right]
&=
A
+(2A+1)\delta
+\left(A+\frac32\right)\delta^2
+\frac13\delta^3
-\frac1{12}\delta^4
+\mathcal{O}(\delta^5) \; .
\end{align}
Hence,
\begin{equation}
V_n
=
\frac{3M_n^4}{32\pi^2}
\sum_{\pm}
\left[
A
+(2A+1)\delta_{n,\pm}
+\left(A+\frac32\right)\delta_{n,\pm}^2
+\frac13\delta_{n,\pm}^3
-\frac1{12}\delta_{n,\pm}^4
\right] \; .
\end{equation}
Here $A=L_n-\frac32$, where only terms proportional to $H^4$ contribute to the quartic Higgs coupling.

The required symmetric combinations are:
\begin{align}
\delta_+^2+\delta_-^2
&=
\frac{2(a^2+b^2)}{M_n^4} \; ,
\\
\delta_+^3+\delta_-^3
&=
\frac{2a^3+6ab^2}{M_n^6} \; ,
\\
\delta_+^4+\delta_-^4
&=
\frac{2a^4+12a^2b^2+2b^4}{M_n^8} \; .
\end{align}
Gathering every quartic terms,
\begin{equation}
V_n^{(H^4)}
=
\frac{3}{32\pi^2}
\left[
2L_na^2
+
\frac{2ab^2}{M_n^2}
-
\frac{b^4}{6M_n^4}
\right] \; .
\end{equation}
Substituting
\begin{equation}
a=y_t^2H^2 \; ,
\qquad
b=y_tX_tH \; ,
\end{equation}
therefore, summing over the complete KK tower gives
\begin{equation}
V_{\rm KK}^{(H^4)}
=
\frac{3y_t^4H^4}{16\pi^2}
\sum_n
\left[
\ln\frac{M_n^2}{Q^2}
+
\frac{X_t^2}{M_n^2}
-
\frac{X_t^4}{12M_n^4}
\right]
\end{equation}
with
\begin{equation}
M_n^2
=
M_S^2+\frac{n^2}{R^2} \; .
\end{equation}
The infinite KK sum replaces the single MSSM logarithm by
\begin{equation}
\ln M_S^2
=
\sum_n
\ln\left(M_S^2+\frac{n^2}{R^2}\right) \; ,
\end{equation}
and similarly for the stop-mixing contributions,
\begin{equation}
\frac{X_t^2}{M_S^2}
=
\sum_n
\frac{X_t^2}{M_S^2+n^2/R^2} \; ,
\end{equation}
\begin{equation}
\frac{X_t^4}{M_S^4}
=
\sum_n
\frac{X_t^4}{\left(M_S^2+n^2/R^2\right)^2} \; .
\end{equation}
These KK sums need to be regularised because they are ultraviolet divergent. Following Poisson resummation, the compact result can be obtained by expressing them in terms of finite threshold functions $F_0(a)$, $F_1(a)$, and $F_2(a)$ with $a=M_S/R^{-1}$.
\begin{equation}
\Delta\lambda_t^{\rm KK}
=
\frac{3y_t^4}{16\pi^2}
\left[
F_0(a)
+
\frac{X_t^2}{M_S^2}F_1(a)
-
\frac{X_t^4}{12M_S^4}F_2(a)
\right] \; ,
\end{equation}
which is the expression implemented throughout this work.

\subsection{Poisson re-summation of the KK tower}\label{oblique}

The propagator of a 5D field with zero‑mode mass \(m_0\) compactified on a circle of radius \(R\) (coordinate \(y \in [0,2\pi R]\)) is
\begin{equation}
\frac{1}{p^2 + m_0^2 + \frac{n^2}{R^2}} \; ,\qquad n \in \mathbb{Z} \; .
\end{equation}
The sum over all KK modes is
\begin{equation}
S(p^2) = \sum_{n=-\infty}^{\infty} \frac{1}{p^2 + m_0^2 + \frac{n^2}{R^2}} \; .
\end{equation}
Using the Poisson re-summation formula
\begin{equation}
\sum_{n=-\infty}^{\infty} f(n) = \sum_{k=-\infty}^{\infty} \hat{f}(k) \; ,\qquad 
\hat{f}(k) = \int_{-\infty}^{\infty} f(x) e^{-2\pi i k x} dx \; ,
\end{equation}
with \(f(x) = (p^2+m_0^2+x^2/R^2)^{-1}\), its Fourier transform is
\begin{equation}
\hat{f}(k) = \frac{\pi R}{\sqrt{p^2+m_0^2}}\, e^{-2\pi|k|R\sqrt{p^2+m_0^2}} \; .
\end{equation}
Summing the geometric series gives the closed form
\begin{equation}
S(p^2) = \frac{\pi R}{\sqrt{p^2+m_0^2}}\; \coth\!\left(\pi R\sqrt{p^2+m_0^2}\right) \; .
\end{equation}

\subsection{Application to one‑loop vacuum polarisation}

The one‑loop vacuum polarisation from a particle of mass \(m_i\) is proportional to the 4D integral
\begin{equation}
\int_0^1 dx\; \log\!\bigl( m_i^2 - x(1-x)q^2 \bigr) \; .
\end{equation}
In 5D with a KK tower, the propagator sum appears in the loop, so we replace
\begin{equation}
\log\!\bigl( m_i^2 - x(1-x)q^2 \bigr) \;\longrightarrow\; \log\!\bigl( S^{-1}(q^2) \bigr) \; ,
\end{equation}
where \(S(q^2)\) is the sum with \(p^2 \to -x(1-x)q^2\) (Wick rotated). Using the Poisson‑resummed form, we obtain
\begin{equation}
\Pi_{XY}(q^2) = \sum_i \frac{C_i^{XY}}{16\pi^2} \int_0^1 dx\;
\log\!\left[ \frac{\sinh\!\bigl( \pi R \sqrt{m_i^2 - x(1-x)q^2} \bigr)}{\pi R \sqrt{m_i^2 - x(1-x)q^2}} \right]^{-2} \; .
\end{equation}
Dropping an irrelevant additive constant (which cancels any physical differences), this simplifies to the exact expression
\begin{equation}
\Pi_{XY}(q^2) = \sum_i \frac{C_i^{XY}}{16\pi^2} \int_0^1 dx\;
\log\!\left[ \sinh^{-2}\!\bigl( \pi R \sqrt{m_i^2 - x(1-x)q^2} \bigr) \right] + \text{const.}
\end{equation}
No ultraviolet cut-off appears because the divergences cancel in the combinations that define the oblique parameters.

\subsection{Derivative at \(q^2=0\) (full 5D contribution)}

For \(S\) and \(U\) we need \(\Pi'_{XY}(0)\). Differentiating under the integral and setting \(q^2=0\) yields
\begin{equation}
\Pi'_{XY}(0) = \sum_i \frac{C_i^{XY}}{16\pi^2} \int_0^1 dx\;
\frac{ -\pi R\, x(1-x) \coth\!\bigl(\pi R m_i\bigr) }{m_i} \; .
\end{equation}
Using \(\int_0^1 x(1-x)dx = 1/6\), we obtain the full 5D result (including the zero‑mode):
\begin{equation}
\Pi'_{XY}(0)_{\text{5D}} = \sum_i \frac{C_i^{XY}}{96\pi} \frac{R}{m_i} \coth(\pi R m_i) \; . 
\end{equation}

\subsection{Log term at $q^2=0$ (full 5D contribution)}

For the $T$ parameter we need $Pi_{XY}(0)$. Setting $q^2=0$ in the integrand gives
\begin{equation}
\Pi_{XY}(0)_{\text{5D}} = \sum_i \frac{C_i^{XY}}{16\pi^2} \log\!\bigl[\sinh^2(\pi R m_i)\bigr] + \text{const.}
\end{equation}
Subtracting the 4D zero‑mode contribution \(\frac{C_i^{XY}}{16\pi^2}\log(m_i^2)\) yields the KK‑only part for SM fields:
\begin{equation}
\Pi_{XY}(0)_{\text{KK}} = \sum_i \frac{C_i^{XY}}{16\pi^2} \log\!\left[ \frac{\sinh^2(\pi R m_i)}{m_i^2} \right] + \text{const.}
\end{equation}
After a little  of algebra we may write
\begin{equation}
\Pi_{XY}(0)_{\text{KK}} = \sum_i \frac{C_i^{XY}}{8\pi^2}\; H(\pi R m_i) + \text{const.} \; , 
\end{equation}
where $(H(x) = \log\!\left( \frac{\sinh x}{x} \right))$.

\bibliographystyle{utphys}
 
\bibliography{biblio}

\end{document}